\documentstyle[eqsecnum,aps]{revtex}
\begin{document}
\draft
\preprint{\vbox{\hbox{DOE/ER/40762-203}\hbox{UMD PP\#00-062}}}
\title{Excited Heavy Baryons and Their Symmetries I: Formalism}
\author{Chi--Keung Chow and Thomas D.~Cohen}
\address{Department of Physics, University of Maryland, College Park, 
20742-4111.}
\date{\today}
\maketitle
\begin{abstract} 
This is the first of two papers to study a new emergent symmetry which 
connects orbitally excited heavy baryons to the ground states in the combined 
heavy quark and large $N_c$ limit.  
The existence of this symmetry is shown in a model-independent way, and 
different possible realizations of the symmetry are discussed.  
It is also proved that this emergent symmetry commutes with the large 
$N_c$ spin-flavor symmetry.  
\end{abstract}

\section{Introduction}

Quantum Chromodynamics is now almost universally accepted as the theory 
which governs strong interaction.  
This theory has been repeatedly tested against experiment, with great 
success.  
Due to its non-abelian nature, however, the QCD coupling gets strong at 
low energy, and the dynamics become nonperturbative and intractable.  
As a result, much of our quantitative understanding of low energy hadron 
properties are based on symmetry considerations.  
The most notable of these schemes is chiral perturbation theory, which is 
based on the fact that, when the light quark masses $m_q\to 0$, the QCD 
Lagrangian is invariant under the chiral symmetry group SU$(n_f)_L\times$ 
SU$(n_f)_R$.  
In the real world, the light masses are not zero; nevertheless chiral 
symmetry survives as an approximate symmetry of QCD, with symmetry breaking 
terms of order $p/\Lambda_\chi$ or $m_\pi/\Lambda_\chi$, where $m_\pi$ is the 
pion mass, $p$ is the scale of external probes, and $\Lambda_\chi$ the chiral 
symmetry breaking scale.  
Despite being just an approximate symmetry, chiral symmetry nevertheless 
provides strong constraints on low energy pion dynamics.  

Important insights into some states in QCD comes from emergent symmetries 
which are {\it not\/} symmetries (not even approximate symmetries) of the QCD 
Lagrangian, but emerge as symmetries of the states in the Hilbert space of an 
effective theory obtained by taking certain limits.  
A famous example of such emergent symmetries is the heavy quark symmetry 
\cite{HQ1,HQ2,HQ3,HQ4,HQ5,HQ6} for heavy hadron states containing a single 
heavy quark with mass $m_Q\gg\Lambda_{\rm QCD}$.
Heavy quark spin symmetry ensures that states related by a heavy quark spin 
flip, like $(B,B^*)$ and $(\Sigma_b,\Sigma_b^*)$, are degenerate.  
Moreover, heavy quark flavor symmetry implies that the brown mucks 
({\it i.e.}, the light degrees of freedom) of heavy hadrons are insensitive 
to the mass or the flavor of the heavy quark.  
This guarantees that the $B\to D^{(*)}$ and $\Lambda_b \to \Lambda_c$ 
semileptonic form factor (which are usually referred as Isgur-Wise form 
factors) are normalized to unity at the point of zero recoil, where the 
initial and final hadrons have the same velocity.  
Such absolute normalizations of form factors have profound implications in 
experimental determination of the CKM matrix element $V_{cb}$.  
The combined heavy quark spin-flavor symmetry is described by the symmetry 
group SU($2n_Q$) where $n_Q$ is the number of heavy flavors, and this 
symmetry is broken by corrections proportional to powers of 
$\Lambda_{\rm QCD}/m_Q$ \cite{HQ1,HQ2,HQ3,HQ4,HQ5,HQ6}.  

Note that heavy quark symmetry is {\it not\/} a symmetry of the QCD 
Lagrangian; if it were, the $B$ and $D$ mesons, related by heavy quark flavor 
symmetry, would be degenerate.  
However, if our interest is restricted to states with a single heavy quark, 
one can perform a spacetime-dependent phase redefinition such that the 
heavy quark mass and spin drop out of the Lagrangian.  
In other words, while heavy quark symmetry is not a symmetry of the QCD 
Lagrangian, it is a symmetry of the Lagrangian of heavy quark effective 
theory, which describes only states with a single heavy quark.  

Another well-known emergent symmetry is the light quark spin-flavor symmetry 
for baryons in the large $N_c$ limit.  
The large $N_c$ limit was first studied by 't Hooft for mesons \cite{LN1} and 
was subsequently extended to baryons by Witten \cite{LN2}.  
They studied how QCD amplitudes involving various numbers of mesons and 
baryons scale with the number of color $N_c$ when $N_c$ is large.  
Gervais and Sakita \cite{LN3,LN4} realized that, for large $N_c$ 
baryons, the spin symmetry SU(2) and flavor symmetry SU($n_f)$ (where $n_f$ is 
the number of light flavors) are combined and enlarged into the spin-flavor 
symmetry group SU($2n_f$).  
(This spin-flavor symmetry was systematically re-analyzed by other groups; 
see Refs.~\cite{LN5,LN6,LN7}.)  
It was shown that the low-lying baryon spectrum in the large $N_c$ limit 
contains a tower of states with $(I,J) = (0,0), (1,1), \dots$ when $N_c$ is 
even, and $({1\over2},{1\over2}), ({3\over2},{3\over2}), \dots$ when $N_c$ is 
odd.  
In the latter case one can identify the $({1\over2},{1\over2})$ state as the 
nucleon, and the $({3\over2},{3\over2})$ as the $\Delta(1232)$ resonance.  
Moreover, it can be shown that the splittings between these tower states are 
of order $1/N_c$.  
As a result, when $N_c\to\infty$, the nucleon, $\Delta$ and all the other 
states in the tower collapse into degeneracy, signifying the emergence of the 
symmetry SU($2n_f$).  
Similarly, in the heavy baryon (baryon with a single heavy quark) sector, 
this spin-flavor symmetry decrees that the $\Sigma_Q^{(*)}$-$\Lambda_Q$ 
splitting vanishes in the large $N_c$ limit.  
Again, note that this light quark spin-flavor symmetry is an emergent 
symmetry in the sense that it is {\it not\/} a symmetry of the QCD Lagrangian, 
but only a symmetry of the QCD Hamiltonian of states with unit baryon number.  

We have recently reported \cite{old} a new emergent symmetry of QCD which 
emerges in the heavy baryon (baryon with a single heavy quark) sector in the 
combined heavy quark and large $N_c$ limit.  
This contracted U(4) symmetry (or more generally U($d+1$) in a theory with $d$ 
spatial dimensions) connects the ground state heavy baryon to some 
of its orbitally excited states, which become degenerate with the ground 
state as $m_Q\to\infty$ and $N_c\to\infty$.  
As a result, static properties such as the axial current couplings and the 
moments of the weak form factors of these orbitally excited states can be 
related to their counterparts of the ground state.  
While many of these results have been discussed before in the literature in 
the context of particular models, they were first presented as 
model-independent symmetry predictions in Ref.~\cite{old}.  

After the publication of Ref.~\cite{old}, we realized that this 
contracted U(4) can be further enlarged into a contracted O(8) symmetry 
(or more generally O($2d+2$) in a theory with $d$ spatial dimensions).  
Moreover, this ``symmetric realization'' is only one of two possible 
realizations of the emergent symmetry.  
In the ``symmetry broken realization'', the symmetry is broken down to 
contracted O(4) $\times$ O(4).  
This paper is the first of two papers where we will report these and other new 
progresses on these emergent symmetries in the combined heavy quark and large 
$N_c$ limits, as well as discuss the results reported in Ref.~\cite{old} in 
more details.  
This paper will focus on the formalism and the different realizations of the 
emergent symmetry, while phenomenological applications and corrections to 
the symmetry will be discussed in the next paper \cite{next}.  

This paper is organized as follow: 
In Sec.~II, we will briefly review the bound state picture, a class of models 
which exhibits this same contracted O($2d+2$) symmetry and which motivates our 
studies.  
While the bound state picture is {\it not\/} logically related to QCD, it 
provides a simple physical picture of the origins of this new symmetry.  
The bound state picture treats the heavy baryon as a bound state of an 
ordinary baryon and a heavy meson and thus is a model.  
However, emergent symmetries are often first recognized in models.  
For example, heavy quark symmetry was first discovered in quark models, while 
the large $N_c$ spin-flavor symmetry was first realized in the Skyrme model.   
Hence it is useful to first consider a model which embodies the correct 
symmetry to get a feeling of the physical picture before launching a formal 
discussion of the symmetry in QCD language.  

After examining the logical foundation of the bound state picture in Sec.~III,
we begin the main task of this paper and study the emergent symmetry in the 
context of QCD.  
In Sec.~IV, we discuss the relative sizes of different contributions 
to the QCD Hamiltonian.  
Then in Sec.~V we introduce the kinematic variables of the bound state 
picture in the context of QCD, and show that many conclusions of the bound 
state picture can be justified in the model independent manner.  
The generators of the emergent symmetry will be formally introduced in Sec.~VI,
and in Sec.~VII and VIII we will focus on the ``symmetric realization'' and 
show that in this case the QCD Hamiltonian is that of a three-dimensional 
simple harmonic oscillator by considering multiple commutation relations in 
the combined heavy quark and large $N_c$ limit.  
Following this is a short discussion in Sec.~IX, while Sec.~X and XI 
will discuss the ``symmetry broken realization'' and the inclusion of spin and 
isospin effects.  
Then the paper concludes with a short preview of the companion paper 
\cite{next}, which is under preparation and will discuss phenomenological 
issues and higher order corrections to the symmetry predictions.  

\section{The bound state picture of a heavy baryon}

The bound state picture 
\cite{genius1,genius2,genius3,CW,others1,others2,others3,others4,others5} 
regards a heavy baryon as the bound state of a heavy meson and a light baryon 
(a baryon without any valence heavy quarks); the latter often treated as a 
chiral soliton, {\it i.e.}, a topologically nontrivial configuration of the 
classical meson fields.  
In particular, the lightest charmed baryon $\Lambda_c$ is regarded as the  
bound state of the heavy mesons $D$ or $D^*$ (which are degenerate in the 
heavy quark limit) and a nucleon.  
In the following, we will focus on the model described in 
Refs.~\cite{genius1,genius2,genius3}, which will be referred to as the simple 
bound state model as it is the simplest model with correct behaviors in the 
heavy quark and large $N_c$ limit.  
However, we emphasize that one can make a similar analysis on other versions 
of the bound state picture, and the symmetry properties should be 
qualitatively the same as long as these models are consistent with heavy quark 
symmetry and obey the usual large $N_c$ counting rules.  

In QCD, a heavy baryon is a complicated bound state, with the quarks 
interacting through strongly coupled gauge dynamics, and with quark-antiquark 
pairs popping in and out of the vacuum, {\it etc.} --- a highly intractable 
problem.  
The bound state picture replaces it (in an {\it ad hoc\/} manner) with the 
much-simpler problem of a two-body bound state.  
Moreover, the problem further simplifies in the heavy quark limit, where the 
heavy meson becomes infinitely massive, and the large $N_c$ limit, where the 
nucleon mass $m_N$ grows like $N_c$.  
For concreteness, we will adopt the prescription (only for this section) that 
the heavy quark limit is taken before the large $N_c$ limit.  \footnote{
In the real world, the heavy meson masses $m_B \sim 5$ GeV, $m_D \sim 1.8$ GeV 
while the nucleon mass $m_N \sim 1$ GeV.  
So as far as heavy baryon kinematics is concerned, the real world is closer to 
the heavy quark limit than the large $N_c$ limit, justifying our ordering of 
the limits.}   
Taking the heavy quark limit first, the reduced mass of the two-body system 
$\mu\sim m_N\sim N_c\to \infty$ in this combined heavy quark--large $N_c$ 
limit.  
As a result, the kinetic term, which is suppressed by $1/\mu$, vanishes, and 
the wave function does not spread but is instead localized at the bottom of 
the potential.  
(When $\mu\to\infty$, the absolute square of the wave function will be a 
Dirac delta distribution at the bottom of the potential.)  
Consequently, a small attraction between the heavy meson and the nucleon is 
sufficient to ensure the existence of a bound state.  

What is the potential ${\cal V}(x)$ between a heavy meson and a nucleon?  
Without resorting to models, we do not know much about the potential except 
the fact that, by the usual large $N_c$ counting rules \cite{LN2}, 
${\cal V}(x)$ is of order $N_c^0$. 
However, let us {\it assume\/} that ${\cal V}(x)$ has a global minimum at the 
origin, {\it i.e.}, the heavy meson sits on the top of (the center of) the 
nucleon.  
In this case, the wave function will be highly localized around the origin,  
and the potential can be approximated by $V(x)= V_0+{1\over2}\kappa\vec x^2$, 
which includes only the first two terms in the Taylor expansion of 
${\cal V}(x)$.  
For the origin to be a global minimum, we need $V_0<0$ and $\kappa>0$.  
In this case, when the bound state is the ground state of the simple harmonic 
oscillator, it is a $\Lambda_Q$.
On the other hand, excited states in the simple harmonic oscillator are
orbitally excited heavy baryons.  
With explicit wave functions, coupling constants and form factors for 
transitions between different states can be calculated in a straightforward 
manner.  

However, it remains to be seen whether the assumptions that $V_0<0$ and 
$\kappa>0$ are justified.  
In the simple bound state model \cite{genius1,genius2,genius3}, the nucleon 
is described as a chiral soliton, {\it i.e.}, a topologically nontrivial 
classical pion configuration.
By using a chiral Lagrangian which is truncated to the leading order, the 
potential energy of a heavy meson in the presence of a classical background 
pion field can be calculated.  
It turns out that indeed $V_0<0$ and $\kappa>0$, and the assumptions are 
verified in this particular model.  
As a result, one can identify different heavy baryons with eigenstates in a 
simple harmonic potential.  
However, this result is clearly model dependent.  

As mentioned before, the bound state picture possesses an emergent symmetry 
which relates the ground state to the excited states.  
This can be seen by making the crucial observation that the excitation 
energy $\omega = \sqrt{\kappa/\mu}$ is small, where $\mu$ is the reduced mass 
of the bound state and $\kappa$ is the spring constant of the simple 
harmonic potential.  
By first taking the heavy quark limit, $\mu = m_N$ (mass of the light 
baryon) scales like $N_c$.  
On the other hand, since the binding potential ${\cal V}(x)$ is of order 
$N_c^0$, the spring constant $\kappa$ --- being its Taylor coefficient --- is 
generically also of order $N_c^0$.  
Hence $\omega$ scales like $N_c^{-1/2}$ and vanishes in the large $N_c$ limit. 
This implies that when $N_c\to\infty$, the whole tower of excited states 
becomes degenerate with the ground state --- a signature of an 
emergent symmetry.  

What is the symmetry group of this emergent symmetry then?  
It has to contain, as a subgroup, the symmetry group of a three-dimensional 
simple harmonic oscillator, namely U(3) generated by $T_{ij} = a_i^\dag a_j$ 
($i,j = 1$, 2, 3) where $a_j$ is the annihilation operator in the $j$-th 
direction.  
These $T_{ij}$'s satisfy the U(3) commutation relations.  
\begin{equation}
[T_{ij},T_{kl}] = \delta_{kj} T_{il} - \delta_{il} T_{kj}.  
\label{com}
\end{equation}
Note that this U(3) group contains the rotational SO(3) subgroup, generated by 
$L_i =-i \epsilon_{ijk} T_{jk}$ with $[L_i,L_j]=i\epsilon_{ijk} L_k$.  
When $N_c \to\infty$ and the excited states become degenerate with the 
ground state, the annihilation and creation operators $a_j$ and $a_i^\dag$ 
($i,j = 1$, 2, 3) also become generators of the emergent symmetry.  
The additional commutation relations are 
\begin{equation} 
[a_j, T_{kl}] = \delta_{kj} a_l, \quad [a_i^\dag, T_{kl}] = -\delta_{il} 
a_k^\dag, \quad [a_j, a_i^\dag] = \delta_{ij} {\bf 1}, 
\end{equation}
where {\bf 1} is the identity operator.  
These sixteen generators $\{T_{ij}, a_l, a_k^\dag, {\bf 1}\}$ form the 
{\it minimal spectrum generating algebra\/} of a three-dimensional harmonic 
oscillator, {\it i.e.}, the smallest algebra which contains the symmetry group 
U(3) and connects all eigenstates of a three-dimensional simple harmonic 
oscillator.  
It is related to the usual U(4) algebra, generated by $T_{ij}$ ($i,j=1$, 2, 
3, 4) satisfying commutation relations (\ref{com}) by the following 
limiting procedure: 
\begin{equation}
a_j = \lim_{R\to\infty} T_{4j}/R, \quad a_i^\dag = \lim_{R\to\infty} T_{i4}/R,
\quad {\bf 1} = \lim_{R\to\infty} T_{44}/R^2.
\end{equation}
Such a limiting procedure is called a group contraction, and hence 
the group generated by $\{T_{ij}, a_l, a_k^\dag, {\bf 1}\}$ is called a 
{\it contracted\/} U(4) group.  \footnote{
This contracted U(4) is different from the contracted SU(4) group of the 
light quark spin-flavor symmetry \cite{LN5}.}

The contracted U(4) minimal spectrum generating algebra can be enlarged to 
contain the extra operators $S_{ij}=a_i a_j$ and $S^\dag_{ij}=a^\dag_i 
a^\dag_j$ ($i,j = 1$, 2, 3).  
The new commutation relations are 
\begin{eqnarray}
[S_{ij},S_{kl}]=[S_{ij},a_l]&=&0, \quad 
[S_{ij},a^\dag_k]=a_i \delta_{jk} + a_j \delta_{ik}, \quad
[S_{ij},T_{kl}]=S_{il} \delta_{jk} + S_{jl} \delta_{ik}, \cr
[S_{ij},S^\dag_{kl}]&=&T_{ik} \delta_{jl} + T_{jk} \delta_{il} 
+ T_{il} \delta_{jk} + T_{jl} \delta_{ik}, 
\end{eqnarray}
and the commutation relations involving $S^\dag_{ij}$ can be obtained 
through Hermitian conjugation.  
As a result, these 28 generators $\{S_{ij}, S^\dag_{ij}, T_{ij}, 
a_l, a_k^\dag, {\bf 1}\}$ form a closed operator algebra, which is actually 
a contracted O(8) algebra.  
This contracted O(8) is generated by the creation and annihilation operators, 
all possible bilinears, as well as the identity operator.  
Note that one cannot further enlarge this operator algebra by including 
trilinears in $a_j$ and $a^\dag_j$; the commutator of two trilinears, for 
example, will be a quadralinear, and the algebra will not close (or will 
contain an infinite number of generators).  
As a result, this contracted O(8) algebra is called {\it maximal spectrum 
generating algebra\/} of the three-dimensional simple harmonic oscillator.  
Again, as $N_c\to\infty$ and $\omega\to0$, the excited states become 
degenerate with the ground state and the contracted O(8) become the symmetry 
group of the bound state picture.  
The relationship between all the algebraic structures discussed above is 
summarized in the following chain: 
\begin{equation}
\matrix{
\hbox{SO(3)} &\subset& \hbox{U(3)} &\subset& \hbox{contracted U(4)} &\subset& 
\hbox{contracted O(8)} \cr
\| & & \| & & \| & & \|\cr
\{L_j\} & & \{T_{ij}\} & & \{T_{ij},a^\dag_i,a_j,{\bf 1}\} & & 
\{S_{ij},S^\dag_{ij}, T_{ij},a^\dag_i,a_j,{\bf 1}\} \cr
& & & & & &\cr
\vbox{\hbox{symmetry group}\hbox{of any}\hbox{central potential}\hbox{}} & &
\vbox{\hbox{symmetry group}\hbox{of 3-D simple}\hbox{harmonic oscillator}
\hbox{}} & &
\vbox{\hbox{minimal spectrum}\hbox{generating algebra,}\hbox{symmetry subgroup}
\hbox{as $\omega\to0$.}} & &
\vbox{\hbox{maximal spectrum}\hbox{generating algebra,}\hbox{symmetry group}
\hbox{as $\omega\to0$.}}}
\end{equation}

We have shown that the contracted O(8) is a symmetry of the bound state 
picture.  
In the more general case of a bound state picture with $d$ spatial dimensions, 
it is clear that the symmetry is described by a similarly contracted O($2d+2$) 
group with a contracted U($d+1$) subgroup.  
While we have been focusing on the simple bound state picture, this symmetry 
is actually a generic feature of all variants of the bound state picture as 
long as the models exhibit heavy quark symmetry as $m_Q\to\infty$, and obey 
the large $N_c$ scaling rules as $N_c\to\infty$.  \footnote{
A note on the literature: as far as the authors can discern, among all the 
literature on the bound state picture for heavy baryon, only the simple bound 
state picture \cite{genius1,genius2,genius3} makes the explicit statement 
that the binding potential is simple harmonic in the combined heavy quark and 
large $N_c$ limit.  
In none of these works on the bound state picture was the spectrum generating 
algebras discussed, nor was the observation that in the combined limit they 
become the symmetry group of an emergent symmetry.  
Both of these points were first explicitly raised in Ref.~\cite{old}.  
On the other hand, the appearance of an emergent symmetry in the bound state 
picture does not depend on the details on the model, as long as the model 
embodies the heavy quark and large $N_c$ symmetry.  
Consequently, the emergent symmetry is an implicit feature of all viable bound 
state models.}  
However, it is not obvious that the physical picture is intuitively 
reasonable.   
This will be addressed in the next section.  

\section{The foundation of the bound state picture}

Questions may be raised about the foundation of the bound state picture on 
several different levels.  
On the {\it technical\/} level, one may question the description of a nucleon 
as a classical pion distribution in the simple bound state model.  
Because we have infinitely many species of mesons in the large $N_c$ 
limit \cite{LN1,LN2}, there is no reason why all other mesons besides 
pions should be ignored.  
This question can be resolved in part by including more light mesons in the 
model.  
This is the motivation behind Ref.~\cite{others3}, where the effects of the 
$\rho$ and $\omega$ vector mesons are included, leading to results which are 
numerically improved at the expense of more parameters and much more 
mathematical complexities.  
Since we are interested in the generic features of the bound state 
picture, we will only remark that including extra meson states does not 
change the physics qualitatively.  
However, in the large $N_c$ limit there is an infinite number of mesons, 
and each meson has infinitely many coupling constants.  

A more serious technical issue of concern is the modeling of the interaction 
between the heavy meson and the classical light meson fields (which make the 
nucleon). 
In the simple bound state picture, the heavy mesons interact with the 
classical background pion configuration through a {\it truncated\/} chiral 
Lagrangian, which is the most general interaction Lagrangian which respects 
chiral symmetry truncated to the leading order of the chiral expansion 
$p/\Lambda_\chi$, where $p$ is the pion momentum.  
While it is justifiable to use this truncated Lagrangian in low momentum 
pion processes (where $p\ll\Lambda_\chi$), there is little justification for 
such truncation here, as the chiral soliton contains pionic modes over a 
wide range of momentum, and in general $p/\Lambda_\chi$ is not a small 
expansion parameter.  
As a result, the truncated Lagrangian is an {\it ad hoc\/} description of the 
interaction between the heavy meson and the classical pion fields.  
The situation is even worse in models where the $\rho$ and $\sigma$ vector 
mesons are included.  
Since the interactions involving these vector mesons are not well-constrained 
by symmetry (unlike pionic interactions, which are severely constrained by 
chiral symmetry), their interactions with heavy mesons are only described by 
phenomenological Lagrangians of an entirely {\it ad hoc\/} nature.  

However, these technological issues are not fundamental and do not alter the 
{\it conceptual\/} issues about the bound state picture.  
For example, it seems likely that, as far as the emergent symmetry is 
concerned, the description of the light baryon as a chiral soliton is not 
essential.  
The essence of the bound state picture is that the heavy baryon can be 
regarded as a bound state with potential ${\cal V}(x)\sim N_c^0$ and reduced 
mass $\mu\sim N_c$.  
The details of the interaction are inessential as far as the symmetry is 
concerned.
This naturally leads us to ask the question whether one can recast the 
analysis in such a form that chiral solitons are not invoked.  
As we will see below, the answer to this question is affirmative.  

A more severe conceptual criticism of the simple bound state model is the use 
of point particle quantum mechanics when both the heavy meson and the nucleon 
are extended objects.  
Assuming that the bound state picture is reasonable, the mean square relative 
displacement of the nucleon from the heavy meson is $3/(2\mu\omega)\sim 
N_c^{-1/2}$, which vanishes as $N_c\to \infty$, while the size of the nucleon 
has a smooth non-zero large $N_c$ limit. 
Hence the heavy meson will be jiggling well inside the nucleon near its 
center, and it is not obvious that point particle quantum mechanics is 
applicable.  

Lastly, the connection of the bound state picture to QCD is obscure.  
To address this philosophical concern, one can only try to reproduce the 
emergent symmetry directly from QCD.  
This is the purpose of both our previous paper \cite{old} and this paper.   
We will see that in a model-independent way, one can show that this contracted 
O(8) symmetry is not only a symmetry of the bound state picture, but in fact 
a symmetry of QCD.  

\section{Dissecting the QCD Hamiltonian for heavy baryons}

Due to the conservation of baryon number and heavy quark number (in the heavy 
quark limit), it is legitimate to restrict our attention to the {\it heavy 
baryon Hilbert space}, {\it i.e.}, the subspace with both heavy quark number 
and baryon number equal to unity.  
In the combined heavy quark and large $N_c$ limit, this subspace is 
well-defined.  
We introduce the small {\it power counting parameter\/}, $\lambda$, to 
quantify the deviation from the combined limit.  
It is defined as:
\begin{equation}
\lambda \sim {\Lambda_{\rm QCD}\over m_Q}, {1\over N_c}.  
\end{equation}
with the ratio $N_c \Lambda_{\rm QCD}/ m_Q$ arbitrary.  
In other words, both $1/m_Q$ and $1/N_c$ corrections are of order $\lambda$, 
while order $\lambda^2$ corrections include those scale like $1/m_Q^2$, 
$1/m_Q N_c$ and $1/N_c^2$, {\it etc.}

In the heavy baryon Hilbert space, it is useful to decompose the QCD 
Hamiltonian $\cal H$ in the following way: 
\begin{equation}
{\cal H} = {\cal H}_Q + {\cal H}_\ell, \qquad \hbox{where} \quad 
{\cal H}_Q = m_Q + \tilde{\cal H}_Q \quad \hbox{and} \quad 
{\cal H}_\ell = m_N + \tilde{\cal H}_\ell.  
\label{h}
\end{equation}
The heavy quark part of the QCD Hamiltonian ${\cal H}_Q$ contains the heavy 
quark mass $m_Q$, as well as the heavy quark kinetic and interaction terms 
denoted by $\tilde {\cal H}_Q$.  
Since $\tilde {\cal H}_Q$ involves only a single quark (namely the heavy 
quark), it at most scales like $N_c^0\sim \lambda^0$.  
(In contrast, both $m_Q$ and $m_N$ are large and of order $\lambda^{-1}$.)  
As we are only interested in states with a single heavy quark, by performing a 
Foldy-Wouthuysen transformation $\tilde {\cal H}_Q$ can be expanded in powers 
of $1/m_Q$ in heavy quark effective theory: 
\begin{equation}
\tilde {\cal H}_Q = gA^0 + {\vec P_Q^2\over 2m_Q} - g{S_Q\cdot B\over 
2m_Q} + {\cal O}(m_Q^{-2}) ,  
\label{hq}
\end{equation}
where $\vec P_Q$ is the three-dimensional heavy quark momentum: 
\begin{equation}
\vec P_Q = \int d^3x \bar h(x) \vec D h(x), 
\label{hp}
\end{equation}
with $h(x)$ being the heavy quark field in heavy quark effective theory, and 
$\vec D$ is the three-dimensional covariant derivative.  
Note that the chromomagnetic term $S_Q\cdot B$ is suppressed by the heavy 
quark mass $m_Q$.  
While the definitions of $m_Q$ and $\vec P_Q$ are ambiguous since the heavy 
quark mass is not uniquely defined, these ambiguities are of order unity 
($\lambda^0$) while the heavy mass and momentum are typically large ($m_Q$, 
$\vec P_Q \sim \lambda^{-1}$).  \footnote{
Both the heavy quark mass $m_Q$ and the heavy quark velocity $v$ are 
well defined up to order $\lambda^0$ ambiguities.  
For a discussion of these ambiguities and their theoretical implications, see
Refs.~\cite{ri1,ri2}.} 
As a result, the relative ambiguities are small and one can rewrite 
Eq.~(\ref{hq}) as 
\begin{equation}
\tilde {\cal H}_Q = gA^0 + {\vec P_Q^2\over 2m_Q} - g{S_Q\cdot B\over 
2m_Q}+ {\cal O}(\lambda^2) .  
\label{hql}
\end{equation}

Similarly, the light part ${\cal H}_\ell$ contains the nucleon mass $m_N$ 
(which is proportional to $N_c$) and $\tilde{\cal H}_\ell$, which 
represents the change in the energy of the brown muck ({\it i.e.,} the light 
degrees of freedom of the heavy baryon) when one of the light quarks is 
replaced by a heavy quark.  
We cannot write down a simple expression for $\tilde{\cal H}_\ell$ as we have 
done for $\tilde {\cal H}_Q$, but it is easy to see that it scales like 
$N_c^0$.  
The reasoning is as follows: 
the interaction energy between any two quarks is of order $N_c^{-1}$ by the 
standard large $N_c$ counting rules \cite{LN1,LN2}.  
Since the replacing of a light quark with a heavy quark in a baryon breaks 
$N_c-1$ light quark--light quark interactions and replaces them with $N_c-1$ 
light quark--heavy quark interactions, we have in the large $N_c$ limit, 
\begin{equation}
\tilde{\cal H}_\ell \sim (\hbox{number of interactions modified}) \times 
(\hbox{change of energy in each interaction}) \sim N_c \times N_c^{-1} 
\sim N_c^0 \sim \lambda^0.  
\end{equation}

The light Hamiltonian $\tilde {\cal H}_\ell$ contains all the dynamics of 
the brown muck as well as its interaction with the heavy quark.  
In general, it can depend not only on its position $\vec x$ and momentum 
$\vec p$ relative to the heavy quark, but also all kinds of internal degrees 
of freedom which correspond to different modes of excitation.  
In comparison, the Hamiltonian ${\cal H}^{\rm bs}$ of a two-particle bound 
state, in general, can be decomposed in the following form: 
\begin{equation}
\tilde {\cal H}^{\rm bs} = {\cal H}_{\rm kin} + {\cal H}_{\rm pot} 
+ {\cal H}_{\rm exc}, 
\label{dh}
\end{equation}
where ${\cal H}_{\rm kin}$ is a kinetic term which depends only on $\vec p$, 
${\cal H}_{\rm pot}$ is a potential term which only depends on $\vec x$, and 
${\cal H}_{\rm exc}$ represents possible internal excitations and commutes 
with both $\vec x$ and $\vec p$.  
The issue becomes whether $\tilde {\cal H}_\ell$ can be recast in this form of 
Eq.~(\ref{dh}).  
This is the question which we will be attempting to answer in the next four  
sections.  

\section{Kinematics and the kinetic energy}

In the previous section, we have decomposed the QCD Hamiltonian, $\cal H$, 
into a heavy part ${\cal H}_Q$ and a light part ${\cal H}_\ell$.  
Our next step will be to perform similar decompositions for the kinematic 
variables; namely, the momentum and position operators.  
We will reproduce the two-body kinematics of the bound state 
picture using QCD operators.  
Recall that our aim is to demonstrate the existence of an emergent symmetry 
in QCD itself.  
In order to achieve this goal in a model-independent manner, one cannot 
simply assume the kinematic variables in the bound state picture are 
well defined.  
Instead we need to construct these kinematic variables from QCD operators 
without reference to any model.  

While the total momentum of any given heavy baryon system $\vec P$ is a 
well-defined quantity, in general there is no unambiguous way to separate 
the momentum into a heavy quark contribution and a brown muck contribution.  
However, in the heavy quark limit, the heavy quark momentum $\vec P_Q$ in 
Eq.~(\ref{hp}) is a well-defined QCD operator (up to corrections of relative 
order $\lambda$), and one can define the brown muck momentum $\vec P_\ell$ as 
$\vec P - \vec P_Q$.  
Lastly, the QCD based position operators of the whole system $\vec X$, of the 
heavy quark $\vec X_Q$, and of the brown muck $\vec X_\ell$ are defined as the 
conjugate operators of the respective momentum operators: 
\begin{equation}
[X_j,P_k] = -i \delta_{ij}, \quad 
[{X_Q}_j,{P_Q}_k] = -i \delta_{ij}, \quad 
[{X_\ell}_j,{P_\ell}_k] = -i \delta_{ij}, \quad 
[x_j,p_k] = -i \delta_{ij},   
\end{equation}
where $X_j$ is the $j$-th component of $\vec X$, {\it etc.}.  
In the last equality, $\vec x$ is defined as the relative position operator 
$\vec X_\ell - \vec X_Q$, and $\vec p$ is its conjugate operator.  
The relationship between these eight operators (four momenta and four 
positions) are summarized in the following diagram: 
\begin{equation}
\matrix{
\vec P&=&\vec P_\ell&+&\vec P_Q& &\vec p\cr
\updownarrow& &\updownarrow& &\updownarrow& &\updownarrow\cr
\vec X& &\vec X_\ell&-&\vec X_Q&=&\vec x\cr}, 
\end{equation}
where vertical arrows represent conjugations.  
Of course these look just like the analogous relations in the bound state 
picture, but recall that the point of introducing these operators 
is to see whether the bound state picture dynamics can be reproduced directly 
from operators in QCD with no model-dependent assumptions.  

By construction, the heavy quark momentum and position operators commute with 
the brown muck counterparts.  
\begin{equation}
[{X_Q}_j,{X_\ell}_k]=[{X_Q}_j,{P_\ell}_k]
=[{P_Q}_j,{X_\ell}_k]=[{P_Q}_j,{P_\ell}_k]=0.
\end{equation}
The center-of-mass position $\vec X$ is an unknown linear combination of 
$\vec X_Q$ and $\vec X_\ell$.  
Similarly, the relative momentum $\vec p$ is an unknown linear combination of 
$\vec P_Q$ and $\vec P_\ell$.  
The operators are defined in such a way that the relative kinematic 
variables commute with the center-of-mass counterparts.  
\begin{equation}
[X_j,x_k]=[X_j,p_k]=[P_j,x_k]=[P_j,p_k]=0, 
\end{equation}
which in turn implies the following linear relations: 
\begin{equation}
\vec X = \hat \alpha \vec X_\ell + \hat \beta \vec X_Q, \qquad
\vec p = \hat \beta \vec P_\ell - \hat \alpha \vec P_Q, 
\label{ab}
\end{equation}
with $\hat \alpha$ and $\hat \beta$ being operators which commute with all the 
momentum and position operators and satisfy $\hat \alpha + \hat \beta = 
{\bf 1}$, the identity operator.  
We emphasize that all of these operators are defined from the QCD operators 
$\vec P$ 
and $\vec P_Q$ through linear combinations and conjugations.  
As a result, all of them are QCD operators.  

\bigskip

What are the operators $\hat \alpha$ and $\hat \beta$?  
To answer this question, one needs to look at the dynamics of the system and 
study the QCD Hamiltonian $\cal H$.  
In particular, we will study the commutators of $\cal H$ with the position 
operators.  

First, consider the commutator $[X_j,{\cal H}]$.  
\begin{equation}
[X_j,{\cal H}] = i \dot X_j = i {P_j\over{\cal H}},  
\end{equation}
where the second equality is from Poincare invariance.  
\footnote{For the reader who does not find the above equality obvious, recall 
that $\cal H$ commutes with $P_j$ (this is one of the defining commutation 
relations of the Poincare group), and hence the rest mass $M$, defined by 
${\cal H}^2 = \sum P_j^2 + M^2$, is Poincare invariant. 
Moreover, being the energy of the whole system at $P_j=0$, $M$ is given by 
the QCD Hamiltonian $\cal H$ in Eq.~(\ref{h}); as a result $M=m_Q+m_N$ 
up to corrections of order $\lambda^0$.  
As a result, $[X_j,{\cal H}] = i{d\over dP_j}{\cal H}=i{d\over dP_j}
(\sum_k P_k^2+M^2)^{1/2} = iP_j (\sum_k P_k^2+M^2)^{-1/2} = 
i {P_j\over{\cal H}}$.  }
Note that the $P_j/{\cal H}$ is well defined as $\cal H$ commutes with $P_j$.  
Moreover, since ${\cal H} = M + \tilde {\cal H}$, where $M=m_Q+m_N\sim
\lambda^{-1}$ while $\tilde {\cal H} \sim \lambda^0$, one can replace 
$1/{\cal H}$ with $1/M$ with relative correction of order $\lambda$.  
As a result, 
\begin{equation}
[X_j,{\cal H}] = i P_j/ M + {\cal O}(\lambda^2), 
\end{equation}
and consequently we have the following double commutators: 
\begin{equation}
[X_k,[X_j,{\cal H}]] = - \delta_{jk}/M , \qquad
[x_k,[X_j,{\cal H}]] = 0 , 
\label{comt}
\end{equation} 
where all ${\cal O}(\lambda^{-2})$ corrections are dropped.  
Note that the first equality is exactly what one expects if one starts with 
the nonrelativistic Hamiltonian, $\vec P^2/2M$. 
As a result, we say that the {\it kinetic mass} of the heavy baryon, defined 
to be the reciprocal of the double commutator with the position operator, is 
$M=m_Q+m_N$ in the combined heavy quark and large $N_c$ limit, with possible 
corrections of order unity.  
These double commutators will be important in the determinations of 
$\hat \alpha$ and $\hat \beta$.  

Next, let us study the commutator $[{X_Q}_j,{\cal H}]$.  
Out of the four contributions to the QCD Hamiltonian $\cal H$ in Eq.~(\ref{h}),
$m_Q$ and $m_N$ are $c$-numbers, and the light operator $\tilde{\cal H}_\ell$ 
commutes with ${X_Q}_j$.    
Thus, 
\begin{equation}
[{X_Q}_j, {\cal H}] = [{X_Q}_j, \tilde{\cal H}_Q] = [{X_Q}_j, 
gA^0 + {\vec P_Q^2\over 2m_Q} - g{S_Q\cdot B\over 2m_Q}] = i{{P_Q}_j\over m_Q} 
+ {\cal O}(\lambda^2),  
\end{equation}
as $A^0$ and $B$ are light operators and commute with $\vec X_Q$.  
On the other hand, by ignoring the ${\cal O}(\lambda^2)$ corrections, one has 
the following double commutators: 
\begin{equation}
[{X_Q}_k,[{X_Q}_j,{\cal H}]] = - \delta_{jk}/m_Q , \qquad
[{X_\ell}_k,[{X_Q}_j,{\cal H}]] = [{X_Q}_k,[{X_\ell}_j,{\cal H}]] = 0,   
\label{comq}
\end{equation}
where the Jacobi identity has been used to show the vanishing of the last 
double commutator.  
Note that the first equality states that, in the heavy quark limit, the 
kinetic mass of the heavy quark is $m_Q$.  

Since $\vec X$ and $\vec x$ are linear combinations of $\vec X_Q$ and 
$\vec X_\ell$ (cf.~Eq.~(\ref{ab})), one can recast the double commutators in 
Eq.~(\ref{comt}) in the following form: 
\begin{eqnarray}
- \delta_{jk}/M =& [X_k,[X_j,{\cal H}]] &= 
{\hat \alpha}^2 [{X_\ell}_k,[{X_\ell}_j,{\cal H}]] + 
{\hat \beta}^2 [{X_Q}_k,[{X_Q}_j,{\cal H}]], \nonumber\\
0 =& [x_k,[X_j,{\cal H}]] &= 
\hat \alpha [{X_\ell}_k,[{X_\ell}_j,{\cal H}]] - 
\hat \beta [{X_Q}_k,[{X_Q}_j,{\cal H}]], 
\end{eqnarray}
where the cross terms vanish by the second formula in Eq.~(\ref{comq}).  
We know $[{X_Q}_k,[{X_Q}_j,{\cal H}]]$ from Eq.~(\ref{comq}) but do {\it 
not\/} know $[{X_\ell}_k,[{X_\ell}_j,{\cal H}]]$ at this stage.  
Canceling the latter, we end up with 
\begin{equation}
- \delta_{jk}/M = (\hat \alpha+ \hat \beta)\hat\beta 
[{X_Q}_k,[{X_Q}_j,{\cal H}]] = - \hat\beta  \delta_{jk}/m_Q, 
\end{equation}
where the relation $\hat \alpha+ \hat \beta = {\bf 1}$ has been used.  
Finally, this implies 
\begin{equation}
\hat\beta = m_Q/M, \qquad \hat\alpha = m_N/M.  
\label{ba}
\end{equation}
Thus the operators $\hat \alpha$ and $\hat \beta$ turn out to be $c$-numbers 
in the combined heavy quark and large $N_c$ limit.  

\bigskip

What do these values of $\hat\alpha$ and $\hat\beta$ tell us about the 
QCD Hamiltonian, $\cal H$?  
First, it is now straightforward to show that the kinetic mass of the brown 
muck is $m_N$.  
\begin{equation}
[{X_\ell}_k,[{X_\ell}_j,{\cal H}]] = - \delta_{jk}/m_N + {\cal O}(\lambda^2). 
\label{coml}
\end{equation}
This, together with Eq.~(\ref{comq}), implies the following decomposition of 
$\cal H$: 
\begin{equation}
{\cal H} = m_Q + m_N + \tilde {\cal H}_{Q, \rm kin} + 
\tilde {\cal H}_{\ell,\rm kin} + \tilde {\cal H}_{\rm pot},   
\end{equation}
where in the combined heavy quark and large $N_c$ limit, 
\begin{equation}
\tilde {\cal H}_{Q, \rm kin} = {\vec P_Q^2 \over 2m_Q} 
+ {\cal O}(\lambda^2), \qquad 
\tilde {\cal H}_{\ell,\rm kin} = {\vec P_\ell^2 \over 2m_N} 
+ {\cal O}(\lambda^2), 
\end{equation}
where the second equality comes from Eq.~(\ref{coml}).  
The potential energy term $\tilde {\cal H}_{\rm pot}\sim\lambda^0$ does not 
depend on any momentum operators (but can and does depend on the position 
operators; see the following section).  
Using Eqs.~(\ref{ba}), these two kinetic terms can be recast in terms of the 
center-of-mass and relative momenta.  
\begin{equation}
\tilde {\cal H}_{\rm kin}=\tilde {\cal H}_{Q, \rm kin} 
+ \tilde {\cal H}_{\ell,\rm kin}= {\vec P^2 \over 2M} 
+ {\vec p^2 \over 2\mu}, 
\label{kin}
\end{equation}
where $M=m_Q+m_N$ is the total mass and $\mu=m_Qm_N/(m_Q+m_N)$ will be 
referred as the {\it reduced mass\/} of the system.  
(Note that both $M$ and $\mu$ are of order $\lambda^{-1}$.)  
The reduced mass $\mu$ can be interpreted as the kinetic mass of the relative 
coordinate.  
Indeed, from the obtained value of $\hat\alpha$ and $\hat\beta$, one can 
verify that 
\begin{equation}
[x_k,[x_j,{\cal H}]] = - \delta_{jk}/\mu + {\cal O}(\lambda^2) 
\label{comr}
\end{equation}
is in agreement with Eq.~(\ref{kin}).  
In other words, in the combined heavy quark and large $N_c$ limit, the kinetic 
terms of the heavy baryon system are those of a nonrelativistic bound state 
of two point particles with $m_Q$ and $m_N$.  

One may wonder what is the point of this whole exercise of reproducing 
elementary two-particle quantum mechanics, with apparently no new results.  
However, remember there is no {\it a priori\/} justification of treating 
a heavy baryon as the bound state of two point particles.  
In particular, the brown muck is not a point particle; it has a 
substantial size and complicated internal structures with the possibility of 
excitations.  
A formalism such as the bound state picture which treats the brown muck as 
though it were a point particle requires justification from QCD.  
Our goal is to demonstrate the existence of an emergent symmetry in QCD 
itself (not merely in a model), we have to work with operators $\vec P$ and 
$\vec P_Q$, which are QCD operators, and construct out of them all other 
kinematic operators.  
That our $\hat \alpha$ and $\hat \beta$ are identical with that in a 
nonrelativistic point particle treatment means that we have succeeded in 
providing a justification of the latter treatment in the combined heavy 
quark and large $N_c$ limit.  

The apparently trivial double commutator in Eq.~(\ref{coml}), which states 
that the kinetic term of the brown muck is nonrelativistic in the combined 
limit, is in fact not completely trivial.  
In the presence of a heavy quark, the kinetic mass of a composite object 
like the brown muck may depend on the relative position of the brown muck to 
the heavy quark.  
What we find instead is a constant kinetic mass $m_N$ in the combined limit.  
Physically this reflects the fact that the system in question is weakly bound; 
the interaction term $\tilde {\cal H}$, which is of order $\lambda^0$, is much 
smaller than the masses of the constituents which are of order 
$\lambda^{-1}$.  
In a weakly bound state, the kinetic mass of the whole system is the sum of 
the kinetic masses of the constituents.  
Since the kinetic mass of the whole system $M$ and that of the heavy quark 
$m_Q$ are position independent, the kinetic mass of the brown muck is also 
independent of its position.  

\section{The generators of the emergent symmetry}

In the previous section, we have analyzed the kinetic terms of the QCD 
Hamiltonian, $\cal H$, by studying its double commutators with the position 
operators.  
One may also want to analyze the potential term $\tilde{\cal H}_{\rm pot}$ 
by studying the double commutators of $\cal H$ with the momentum operators.  
Unfortunately, this strategy actually provides very limited information.  

As Poincare invariance demands that $[P_j,{\cal H}]=0$, it immediately follows 
that $[{P_\ell}_j,{\cal H}]=-[{P_Q}_j,{\cal H}]$.  
Moreover, 
\begin{equation}
[p_j,{\cal H}]=\hat\beta[{P_\ell}_j,{\cal H}] - \hat\alpha[{P_Q}_j,{\cal H}] 
= (\hat\alpha+\hat\beta)[{P_\ell}_j,{\cal H}] = [{P_\ell}_j,{\cal H}], 
\end{equation}
where again we have used $\hat\alpha+\hat\beta = {\bf 1}$.  
This reflects the simple observation that $\tilde {\cal H}_{\rm pot}$ can 
depend on the relative position $\vec x$ but not the center-of-mass position 
of the whole system $\vec X$.  

Recall that we have made the following decomposition: ${\cal H} = m_Q + m_N 
+ \tilde {\cal H}_{\rm kin} + \tilde {\cal H}_{\rm pot}$, with the last two 
terms both being of order unity or less.  
Both $m_Q$ and $m_N$ are $c$-numbers and commute with any operator, and 
it is easy to see that $\tilde {\cal H}_{\rm kin}$ commutes with all momentum 
operators from its expression in Eq.~(\ref{kin}), which does not depend on 
any of the position operators.  
As a result, $\tilde {\cal H}_{\rm pot}$ is the only term which may have 
non-vanishing commutators with the momentum operators.  
The commutators of interest are $[p_j,\tilde {\cal H}_{\rm pot}]$, 
$[p_k,[p_j,\tilde {\cal H}_{\rm pot}]]$, 
$[p_k,[p_j,[p_i,\tilde {\cal H}_{\rm pot}]]]$, {\it etc.}  
We can say very little about these multiple commutators except 
that they are all (at most) of the same order as $\tilde {\cal H}_{\rm pot}$, 
{\it i.e.}, of order unity.  

Of particular interest is the double commutator, whose significance lies in 
the following definition of the {\it spring operator\/} $\hat\kappa$: 
\begin{equation}
\hat\kappa \delta_{jk} = - [p_k,[p_j,{\cal H}]] 
= - [p_k,[p_j,\tilde {\cal H}_{\rm pot}]].  
\label{so}
\end{equation} 
Let $|G\rangle$ be the ground state of the QCD Hamiltonian $\cal H$ {\it in 
the heavy baryon Hilbert space\/}, and $E_0$ be its mass, satisfying 
$({\cal H} - E_0)|G\rangle = 0$.  
The {\it spring constant\/} $\kappa$ is then defined as 
$\langle G|\hat\kappa|G\rangle$.  
Since \footnote{In the following two equations, the repeated index $j$ is 
{\it not\/} being summed over.}
\begin{equation}
\hat\kappa = - [p_j,[p_j,{\cal H}]] = - [p_j,[p_j,({\cal H}-E_0)]] = 
2p_j ({\cal H}-E_0) p_j - p_j p_j ({\cal H}-E_0) - ({\cal H}-E_0)p_j p_j, 
\end{equation}
it is easy to see that $\kappa$ is positive by inserting a complete set of 
states $\{|n\rangle\}$. 
\begin{equation}
\kappa = 2 \langle G|p_j ({\cal H}-E_0) p_j |G\rangle = 2 \sum_n |\langle n|
p_j|G\rangle|^2 (E_n-E_0) > 0.  
\end{equation}

Now we are ready to define the generators of the emergent symmetry; namely, 
the creation and annihilation operators.  
\begin{equation}
\vec a = \sqrt{\mu\omega\over2}\vec x + i\sqrt{1\over2\mu\omega}\vec p, \qquad 
\vec a^\dag = \sqrt{\mu\omega\over2}\vec x - i\sqrt{1\over2\mu\omega}\vec p, 
\end{equation}
where $\mu = m_Qm_N/(m_Q+m_N)$ is the reduced mass and $\omega = 
\sqrt{\kappa/\mu}$ will be referred to as the {\it natural frequency\/} of the 
heavy baryon.  
With $\kappa$ of order unity and $\mu\sim\lambda^{-1}$ in the combined heavy 
quark and large $N_c$ limit, $\omega$ vanishes --- an important result which 
does not depend on the order in which the two limits are taken.  
If the heavy quark limit is taken first, $\mu \sim N_c$ and $\omega \sim 
N_c^{-1/2} \to 0$ as $N_c\to\infty$.  
On the other hand, if the large $N_c$ limit is taken first, $\mu \sim m_Q$ 
and $\omega \sim m_Q^{-1/2} \to 0$ as $m_Q\to\infty$.  

The natural frequency $\omega$ plays a central role in heavy baryon 
dynamics in the combined heavy quark and large $N_c$ limit.  
As we will see below, $\omega$ is the coefficient of the only term in the 
QCD Hamiltonian which breaks the emergent symmetry at leading order of 
$\lambda$.  
As a result, all the physical properties (masses, coupling constants, form 
factors, {\it etc.}) of the low-lying baryons can be expressed in terms of 
$\omega$.  
Alternatively, if one can determine the value of $\omega$ by measuring 
some physical observable which depends on $\omega$ ({\it e.g.}, the mass of 
the first excited heavy baryon), then one can predict the values of many other 
physical observables.  

\section{Constraints on the QCD Hamiltonian}

Let us recall that our analysis is based on $\lambda$ counting of quantities 
describing heavy baryon dynamics in the combined limit.  
The reduced mass can be expressed as the double commutator in Eq.~(\ref{comr}):
\begin{mathletters}
\begin{eqnarray}
\delta_{jk}/\mu &= - [x_k,[x_j,{\cal H}]] \sim& \lambda, \\
\noalign{\smallskip

\noindent where higher order terms in $\lambda$ are dropped.  
Similarly, the spring constant $\kappa$ is the ground state expectation value 
of the double commutator in Eq.~(\ref{so}):

\bigskip}
\delta_{jk} \hat\kappa &= - [p_k,[p_j,{\cal H}]] \sim& \lambda^0, \qquad 
\kappa = \langle G|\hat\kappa|G\rangle.
\end{eqnarray}
\label{dc}
\end{mathletters}
As a result, the natural frequency $\omega=\sqrt{\kappa/\mu}
\sim\lambda^{1/2}$.  
This is a notable feature: recall that the expansion in $\lambda$ embodies 
the expansions in both $\Lambda_{\rm QCD}/m_Q$ and $1/N_c$.  
We rarely encounter situations in which fractional powers of 
$\Lambda_{\rm QCD}/m_Q$ or $1/N_c$ arise for direct physical observables.  
Here, however, we have found that powers like $\lambda^{1/2}$ do arise 
naturally.  
In fact, we will see that the natural expansion parameter will be 
$\lambda^{1/2}$ instead of $\lambda$.  
This ultimately reflects the interplay of two heavy scales (namely $m_Q$ and 
$m_N$).  

The double commutation relations in Eqs.~(\ref{dc}) constrain the possible 
forms of the QCD Hamiltonian $\cal H$.  
Note, however, that both double commutation relations are satisfied by 
replacing $\cal H$ with ${\cal H}_{\rm SHO}$, the Hamiltonian of the bound 
state picture, which is just the simple harmonic oscillator.  
\begin{equation}
{\cal H}_{\rm SHO}= {\vec p^2 \over 2\mu} + {\kappa \vec x^2\over2} 
- {3\omega\over 2} = \omega \vec a_j^\dag \cdot \vec a_j.  
\end{equation}
Moreover, the contracted O(8) symmetry mentioned above is precisely the 
maximal spectrum generating algebra of ${\cal H}_{\rm SHO}$ and becomes an 
emergent symmetry as $\omega\to 0$ as $\lambda\to 0$ in the combined limit.  
On the other hand, to demonstrate that this contracted O(8) is a symmetry of 
QCD in this limit, one needs to show that the generators of the contracted 
O(8) commute with the QCD Hamiltonian $\cal H$, or equivalently, show that 
\begin{equation}
{\cal H} = {\cal H}_{\rm SHO}+{\cal H}_{\rm exc}+\dots, 
\label{hd}
\end{equation}
where ${\cal H}_{\rm exc}$ commutes with $\vec a$ and $\vec a^\dag$ in the 
combined limit, {\it i.e.}, $[a_j,{\cal H}_{\rm exc}]=[a_j^\dag,
{\cal H}_{\rm exc}]=0$, and represents the possibility of internal excitations 
of the brown muck.  
On the other hand, the ellipses are possible corrections to the simple 
harmonic Hamiltonian ${\cal H}_{\rm SHO}$ and should be negligible in 
comparison to ${\cal H}_{\rm SHO}$ in the combined limit.  
In other words, we need to show that, relative to ${\cal H}_{\rm SHO}$, 
these correction terms are suppressed by powers of $\lambda$, and hence can be 
dropped in the combined limit.  

Before we embark the power counting of $\cal H$, we need to clarify the 
counting of powers for the kinematic variables.  
For example, consider the simple harmonic Hamiltonian ${\cal H}_{\rm SHO} = 
\omega \vec a^\dag \cdot \vec a$, which {\it apparently\/} is of order 
$\lambda^{1/2}$ as $\omega\sim\lambda^{1/2}$. 
However, ${\cal H}_{\rm SHO}$ can also be written as ${\vec p^2\over 2\mu} 
+ {\kappa \vec x^2\over 2} - {3\omega\over 2}$, 
where the kinetic term, being suppressed by $1/\mu$, is {\it apparently\/} of 
order $\lambda$, while the potential term, with coefficient 
$\kappa\sim\lambda^0$, is {\it apparently\/} of order unity.  
The origins of this apparent discrepancy lie in the relationship between the 
operators $(\vec x,\vec p)$ and $(\vec a,\vec a^\dag)$.  
\begin{equation}
\vec x = \sqrt{1\over 2\mu\omega} (\vec a + \vec a^\dag), \qquad
\vec p = -i \sqrt{\mu\omega\over 2} (\vec a - \vec a^\dag).  
\end{equation}
Since $\mu\omega\sim\lambda^{-1/2}$, one cannot simultaneously set $\vec x$, 
$\vec p$, $\vec a$ and $\vec a^\dag$ to the same order in $\lambda$.  

In the following, we will make the prescription that $\vec a, \vec a^\dag \sim 
\lambda^0$, which implies $\vec x\sim\lambda^{1/4}$ and 
$\vec p\sim\lambda^{-1/4}$, and show that it is self-consistent.    
As we will see later in this paper, this prescription will lead to the 
``symmetric realization'' of the emergent symmetry with a contracted O(8) 
symmetry group.  
We can make the following {\it a posteriori\/} justification for this 
``symmetric prescription''.  
Instead of studying the power counting of the {\it operators\/} $\vec x$, 
$\vec p$, $\vec a$ and $\vec a^\dag$, one can study the power counting of 
the {\it matrix elements\/} of $\vec x$, $\vec p$, $\vec a$ and $\vec a^\dag$ 
between the low-lying states of $\cal H$.  
Power counting on matrix elements, which are $c$-numbers, is free of the 
aforementioned ambiguities.  
If the low-lying states of $\cal H$ are simple harmonic states as suggested 
by the bound state picture, then the matrix elements of $\vec a$ and 
$\vec a^\dag$ are indeed of order unity, while the matrix elements of $\vec x$ 
and $\vec p$ are not.  
As the subsequent discussion verifies this picture, we will have justified our 
prescription {\it a posteriori}.  

Now we are ready to show that in the combined limit $\cal H$ has the form 
presented in Eq.~(\ref{hd}). 
We will achieve this in three steps.  
In the remainder of this section, we will verify the fact that all possible 
triple commutators of $\cal H$ with $\vec a$ and $\vec a^\dag$ vanish in the 
combined limit.  
More specifically, we will show that these triple commutators go to zero 
more quickly than ${\cal H}_{\rm SHO}$, which scales like $\lambda^{1/2}$.  
Then in the following section, we will show how the vanishings of these 
commutators imply that $\cal H$ can be at most bilinear in $\vec a$ and 
$\vec a^\dag$:
\begin{equation}
{\cal H}= \hat C \vec a^\dag \cdot \vec a + \hat D \vec a \cdot \vec a 
+ \hat D^\dag \vec a^\dag \cdot \vec a^\dag + {\cal H}_{\rm exc}, 
\label{form}
\end{equation}
where ${\cal H}_{\rm exc}$ commutes with both $\vec a$ and $\vec a^\dag$.  
Lastly, $\hat C$ and $\hat D$ can be determined from the double commutation 
relations in Eqs.~(\ref{dc}).  

\bigskip

The relevant triple commutators of $\cal H$ with $\vec a$ and $\vec a^\dag$ 
are the following operators: 
\begin{eqnarray}
t^{(0)}&=[a_i,[a_j,[a_k,{\cal H}]]], \qquad 
t^{(1)}&=[a^\dag_i,[a_j,[a_k,{\cal H}]]], \nonumber\\
t^{(2)}&=[a^\dag_i,[a^\dag_j,[a_k,{\cal H}]]], \qquad 
t^{(3)}&=[a^\dag_i,[a^\dag_j,[a^\dag_k,{\cal H}]]].  
\end{eqnarray}
Using the Jacobi identity and the commutators of $a_j$ and $a_j^\dag$, it is 
easy to show that the values of these $t^{(a)}$ triple commutators do not 
depend on the ordering of the $a_j$'s and $a^\dag_j$'s.  
For example, 
\begin{eqnarray}
[a^\dag_i,[a_j,[a_k,{\cal O}]]] 
&= -[a_j,[[a_k,{\cal O}],a^\dag_i]] - [[a_k,{\cal O}],[a^\dag_i,a_j]] 
=& [a_j,[a^\dag_i,[a_k,{\cal O}]]] \nonumber\\
&= -[a_j,[a_k,[{\cal O},a^\dag_i]]] - [a_j,[{\cal O},[a^\dag_i,a_k]]]
=& [a_j,[a_k,[a^\dag_i,{\cal O}]]].  
\end{eqnarray}
Since $t^{(0)} = -\big(t^{(3)}\big)^\dag$ and $t^{(1)} = 
-\big(t^{(2)}\big)^\dag$, it suffices to show that $t^{(0)}=t^{(1)}=0$ in the 
combined limit.  

Since $\vec a$ and $\vec a^\dag$ are linear combinations of $\vec x$ and 
$\vec p$, $t^{(0)}$ and $t^{(1)}$ can be expressed as linear combinations of 
the triple commutators of $\cal H$ with $\vec x$ and $\vec p$:  
\begin{equation}
t^{(0)} = 2^{-3/2} (T^{(3)} + 3i T^{(2)} - 3 T^{(1)} - i T^{(0)}), \qquad 
t^{(1)} = 2^{-3/2} (T^{(3)} + i T^{(2)} + T^{(1)} + i T^{(0)}),
\end{equation}
where 
\begin{eqnarray}
T^{(0)} &= (\mu\omega)^{-3/2} [p_i,[p_j,[p_k,{\cal H}]]],\qquad
T^{(1)} &= (\mu\omega)^{-1/2} [p_i,[p_j,[x_k,{\cal H}]]],\nonumber\\
T^{(2)} &= (\mu\omega)^{1/2} [p_i,[x_j,[x_k,{\cal H}]]],\qquad\;\;
T^{(3)} &= (\mu\omega)^{3/2} [x_i,[x_j,[x_k,{\cal H}]]].
\end{eqnarray}
Again note that the values of these $T^{(a)}$ triple commutators do not depend 
on the ordering of the $x_j$'s and $p_j$'s.  
We have shown in Sec.~VI that the triple $p$ commutator is at most of order 
unity, and hence $T^{(0)}\sim {\cal O}(\lambda^{3/4})$.  
All of the other three triple commutators are also small as  
$[x_k,{\cal H}] = ip_k/\mu + {\cal O}(\lambda^2)$.  
The first term gets killed by the following commutations and does not 
contribute to the triple commutators.  
So $T^{(1)}\sim \lambda^{9/4}$, $T^{(2)}\sim \lambda^{7/4}$ and $T^{(3)}\sim 
\lambda^{5/4}$ --- all vanish faster than $T^{(0)}$.  
As a result, both $t^{(0)}$ and $t^{(1)}$ vanish at least as fast as 
$T^{(0)}\sim\lambda^{3/4}$, and are negligible when compared to 
${\cal H}_{\rm SHO}\sim\lambda^{1/2}$ in the combined limit.  

\section{The QCD Hamiltonian in the combined limit}

In the previous section, we have verified that all triple commutators 
$t^{(a)}$ vanish faster than ${\cal H}_{\rm SHO}\sim\lambda^{1/2}$ in the 
combined limit.  
In other words, if we only keep terms up to those of order $\lambda^{1/2}$, 
all these triple commutators are zero.  
\begin{equation}
[a_i,[a_j,[a_k,{\cal H}]]] = [a^\dag_i,[a_j,[a_k,{\cal H}]]] = 
[a^\dag_i,[a^\dag_j,[a_k,{\cal H}]]]=[a^\dag_i,[a^\dag_j,[a^\dag_k,{\cal H}]]]
\leq {\cal O}(\lambda^{-3/4}).  
\end{equation}

Now the vanishings of these triple commutators at order $\lambda^{1/2}$ imply 
that to this order the double commutators commute with $\vec a$ and 
$\vec a^\dag$:
\begin{equation}
[a_j,[a^\dag_k,{\cal H}]] = [a^\dag_j,[a_k,{\cal H}]] = - \hat C \delta_{jk}, 
\qquad [a^\dag_j,[a^\dag_k,{\cal H}]] = 2 \hat D \delta_{jk}, \quad 
[a_j,[a_k,{\cal H}]] = 2 \hat D^\dag \delta_{jk}, 
\label{step1}
\end{equation}
where the operators $\hat C$ and $\hat D$ commute with both $\vec a$ and 
$\vec a^\dag$.  
Note that $\hat C$ is hermitian while $\hat D$ is in general non-hermitian.  
The above relations can be recast as: 
\begin{equation}
[a_j, \hat B] = 0, \quad [a^\dag_j, \hat B] = 0, \qquad \hbox{where} \quad 
\hat B = [a^\dag_k,{\cal H}]+\hat C a^\dag_k+2\hat D a_k.  
\end{equation}
However, parity invariance of $\cal H$ implies that there does not exist any 
parity odd operator which commutes with both $\vec a$ and $\vec a^\dag$.  
So $\hat B$ vanishes and 
\begin{mathletters}
\begin{equation}
[a^\dag_k,{\cal H}]+\hat C a^\dag_k+2\hat D a_k = 0. 
\end{equation}
Similarly, one has 
\begin{equation}
[a_k,{\cal H}]-\hat C a_k-2\hat D^\dag a^\dag_k = 0.
\end{equation}
\label{step2}
\end{mathletters}
Notice that, while in Eqs.~(\ref{step1}) we expressed the double commutators 
of $\cal H$ in term of $\hat C$ and $\hat D$, in Eqs.~(\ref{step2}) we 
managed to express the single commutators of $\cal H$ in term of $\hat C$ and 
$\hat D$.  
We can make one more step and express $\cal H$ itself in terms of $\hat C$ and 
$\hat D$ by noting that Eqs.~(\ref{step2}) imply
\begin{equation}
[a_j, \hat A] = 0, \quad [a^\dag_j, \hat A] = 0, \qquad \hbox{where} \quad 
\hat A = {\cal H}- \hat C \vec a^\dag \cdot \vec a + \hat D \vec a \cdot 
\vec a + \hat D^\dag \vec a^\dag \cdot \vec a^\dag,  
\end{equation}
where $\hat A$ commutes with both $\vec a$ and $\vec a^\dag$.  
Lastly, by renaming $\hat A$ as ${\cal H}_{\rm exc}$, we recover 
Eq.~(\ref{form}):  
\begin{equation}
{\cal H}= \hat C \vec a^\dag \cdot \vec a + \hat D \vec a \cdot \vec a 
+ \hat D^\dag \vec a^\dag \cdot \vec a^\dag + {\cal H}_{\rm exc}.   
\end{equation}

We have succeeded in showing that the vanishings of the triple commutations 
of $\cal H$ imply that $\cal H$ is at most a bilinear in $\vec a$ and 
$\vec a^\dag$. 
\footnote{This is actually a special case of the following general result. 
Let $\cal O$ be an arbitrary operator with vanishing $m$-th multiple 
commutators with $\vec a$ and $\vec a^\dag$.  
Then one can prove by induction that $\cal O$ is at most $(m-1)$-linear in 
$\vec a$ and $\vec a^\dag$. }
While the above derivation looks rather complicated, the essence of the 
statement is very intuitive: 
if $\cal H$ contains, for instance, trilinear terms in $\vec a$ and 
$\vec a^\dag$, the triple commutators will read off the coefficients of these 
trilinear terms and hence will not vanish.  
What we have achieved, through the derivation above, is to realize this 
intuition in a rigorous manner.  

\bigskip

We have shown that, up to order $\lambda^{1/2}$, $\cal H$ is a bilinear in 
$\vec a$ and $\vec a^\dag$.  
Now we will make the final step in this demonstration and determine the form 
of the operators $\hat C$ and $\hat D$.  
Clearly if $\hat C = \omega$ and $\hat D = 0$, $\cal H$ will simply be the sum 
of the simple harmonic Hamiltonian ${\cal H}_{\rm SHO}$ and a possible 
excitation term ${\cal H}_{\rm exc}$ which commutes with $\vec a$ and 
$\vec a^\dag$. 
Our goal is to show that these are indeed the values of $\hat C$ and $\hat D$. 

One can re-express $\cal H$ in terms of $\vec x$ and $\vec p$: 
\begin{equation}
{\cal H} = {(\hat C - \hat D_+)\over2\mu\omega} \vec p^2 
+ {\mu\omega(\hat C + \hat D_+)\over2} \vec x^2 - {3\hat C\over 2} 
- {\hat D_-\over 2} (\vec x \cdot \vec p + \vec p \cdot \vec x) 
+ {\cal H}_{\rm exc}, 
\label{almost}
\end{equation}
where $\hat D_+ = \hat D + \hat D^\dag$ and $i\hat D_- = \hat D - \hat D^\dag$.
Then $\hat C$, $\hat D_+$ and $\hat D_-$ can be deduced by using the double 
commutation relations Eqs.~(\ref{dc}): 
\begin{eqnarray}
[p_j,[x_k,{\cal H}]] =&  0 \qquad \quad\,\Rightarrow \qquad 
\hat D_- & = 0, \nonumber\cr
\noalign{\smallskip}
[x_j,[x_k,{\cal H}]] =&  -\delta_{jk}/\mu \quad \Rightarrow \quad 
\hat C - \hat D_+ & = \omega, \cr
\noalign{\smallskip}
[p_j,[p_k,{\cal H}]] =&  -\delta_{jk}\hat \kappa \quad\;\; \Rightarrow \quad
\hat C + \hat D_+ & = \hat\kappa/\mu\omega = \omega \hat\kappa/\kappa.   
\end{eqnarray}
The last two equalities lead to:
\begin{equation}
\hat C = \omega (\hat\kappa/\kappa+1)/2, \qquad 
\hat D_+ = \omega (\hat\kappa/\kappa-1)/2,   
\end{equation}
where the spring constant $\kappa$ is the ground state expectation value of 
the spring operator $\hat\kappa$ introduced in Eq.~(\ref{so}).  
As a result, {\it when acting on the heavy baryon ground state\/}, 
$\hat\kappa/\kappa = 1$, which in turn implies $\hat C=\omega$ and 
$\hat D_+=0$.  
For a general heavy baryon state (not necessarily the ground state), however, 
$\hat\kappa$ is {\it not\/} identical to its ground state expectation value 
$\kappa$.  
However, it is true not merely for the ground state, but also for states in 
the {\it ground state band}, which is the subspace spanned by states of the 
form $(a_x^\dag)^{n_x} (a_y^\dag)^{n_y} (a_z^\dag)^{n_z}|G\rangle$.  
We therefore conclude that, in the ground state band, $\hat C=\omega$, 
$\hat D_+=0$, and $\cal H$ has the simple harmonic form:  
\begin{equation}
{\cal H} = {\vec p^2\over 2\mu} + {\kappa \vec x^2\over2} 
- {3\omega\over 2} + {\cal H}_{\rm exc} + {\cal O}(\lambda)
= \omega \vec a^\dag \cdot \vec a + {\cal H}_{\rm exc} + {\cal O}(\lambda).  
\label{result}
\end{equation}
This Hamiltonian clearly reduces to a three-dimensional simple harmonic 
oscillator with reduced mass $\mu$, spring constant $\kappa$, and hence 
natural frequency $\omega$.  
The maximal spectrum generating algebra of this Hamiltonian is a contracted 
O(8) algebra, and in the combined limit when $\omega\sim\lambda^{1/2}\to0$, 
the contracted O(8) becomes an emergent symmetry.  
Again, we emphasize that this emergent symmetry is a symmetry of QCD in 
the heavy baryon sector --- not only that of the bound state picture or any 
other models.  

\section{Review and Discussion}

In summary, we have demonstrated that the contracted O(8) symmetry seen in the 
bound state picture is in fact a symmetry of QCD.  
Near the combined limit there exists a band of low-lying heavy baryons, 
labeled by $(n_x, n_y, n_z)$, the number of excitation quanta in the $x$, $y$ 
and $z$ directions (or alternatively $(N,L,L_z)$, where $N=n_x+n_y+n_z$ is 
the total number of excitation quanta, $L$ the orbital angular momentum and 
$L_z$ its $z$-component).  
For each state the excitation energy is $(n_x+n_y+n_z)\omega=N\omega$. 
As $\lambda\to0$ in the combined limit, $\omega\to0$ and the entire band 
become degenerate.  
Our discussion can be generalized in a straightforward manner to the case 
with $d$ spatial dimensions, with an emergent contracted O($2d+2$) symmetry 
group.  

While the demonstration was rather long, the basic idea is very simple.  
We started by constructing the kinematic variables $x_j$ and $p_j$, which are 
not {\it a priori\/} well defined in QCD (Sec.~V).  
Since our goal is to study the symmetry in QCD itself, we cannot merely 
assume that the kinematic variables are well defined (as in models), but 
need to show that they are legitimate QCD operators.  
After defining these kinematic variables in QCD, we showed that they behave 
like the quantum mechanical kinematic operators for bound states of two point 
particles. 
Thus they can be used in a ``quantum mechanics-like'' framework to describe 
the dynamics in the heavy baryon sector.  
With the creation and annihilation operators constructed from $x_j$ and $p_j$
(Sec.~VI), we show, by considering triple commutators (Sec.~VII), that the QCD 
Hamiltonian $\cal H$ is a bilinear of these creation and annihilation 
operators up to a certain order in the $\lambda$ expansion (Sec.~VIII).  
Lastly, the ``coefficients'' (which are formally operators) of the bilinear 
terms in $\cal H$ are fixed by considering double commutators.  

It may seem strange that a symmetry of QCD is only applicable 
to a certain subspace (namely, the ground state band of the heavy baryon 
subspace) of the whole QCD Hilbert space.  
This, however, is a typical feature for emergent symmetries.  
The heavy quark symmetry is only applicable to states containing a heavy 
quark \cite{HQ1,HQ2,HQ3,HQ4,HQ5,HQ6}, and the light quark spin-flavor symmetry 
in the large $N_c$ limit is only relevant for baryonic states 
\cite{LN3,LN4,LN5,LN6,LN7}.  
While the symmetries of the QCD Lagrangian is applicable to all states in 
the QCD Hilbert space, emergent symmetries are {\it not\/} symmetries of 
the QCD Lagrangian and may be applicable only to particular subspaces.  

We will end this section by briefly returning to the bound state picture 
and discuss some intricate issues on its relationship to our formalism. 
We noted in Sec.~III that there were conceptual problems associated
with the bound state picture, particularly in regard to treating the brown
muck as though it were a point particle.  
The possible problem was that the characteristic size of the brown muck 
distribution is $L_q^2 \sim \Lambda_{\rm QCD}\sim\lambda^0$ while the bound 
state wave function had a typical size of $L_{\rm wf}^2\sim (\mu\omega)^{1/2} 
\sim \lambda^{1/2}$ which is characteristically narrower in position space.  
In spite of this concept we have shown that as far as kinematics and 
symmetries are concerned, the point particle description correctly reproduces 
the QCD result.
It turns out that the comparison between the characteristic size of
the brown muck and the scale of the wave function is not the appropriate
comparison.  
In fact, there are three distance scales in this problem: 
If the brown muck is to be approximated by a point particle, the point 
particle should be located at the center-of-mass of the brown muck in order 
to reproduce the correct kinematics.  
According to Ref.~\cite{LN2}, the brown muck can be studied under the Hartree 
picture, which becomes exact in the large $N_c$ limit.  \footnote{
Actually Ref.~\cite{LN2} used the Hartree picture to study a baryon, not 
the brown muck of a heavy baryon.  
However, the difference between a brown muck of a heavy baryon, with $N_c-1$ 
light quarks, and a light baryon with $N_c$ light quarks, becomes negligible 
in the large $N_c$ limit. 
(Of course, the brown muck is different from a baryon as the former is not a 
color singlet, but the extra color charge is neutralized by the heavy quark.)}
In the Hartree picture, one can easily see that the center-of-mass of the 
brown muck is much better localized than each individual quark.  
The size of the wave function of each individual quark $L_q$ is comparable to 
the size of the whole brown muck, which is of order unity.  
On the other hand, the fluctuation of the position of center-of-mass $L_{CM}$ 
of the brown muck is smaller by a factor of $\sqrt N_c$ by the central limit 
theorem.  
As a result, $L_{CM}^2 \sim L_q^2/N_c \sim \lambda$, which is much smaller 
than the typical spread of the wave function.  
The different scales form the following hierarchy: 
\begin{equation}
L_q^2 \sim \lambda^0 \quad \gg \quad L_{\rm wf}^2 \sim \lambda^{1/2} 
\quad \gg \quad L_{\rm CM}^2 \sim \lambda.  
\end{equation}
In other words, the bound state wave function cannot resolve the fluctuation 
of the center-of-mass of the brown muck.  
This provides an intuitive explanation why, despite its huge size, the 
brown muck can be approximated by a point particle without drastically 
altering the kinematics and the symmetries of the system.  

\section{The symmetry broken realization}
  
Recall that our demonstration of the simple harmonic form of the QCD 
Hamiltonian $\cal H$ depends on the symmetric prescription for the 
$\lambda$ power counting introduced in Sec.~VII, that the creation and 
annihilation operators should be counted as order unity in $\cal H$.  
This prescription is in turn {\it a posteriori\/} justified by the fact 
that the matrix element of $a_j$ and $a^\dag_j$ between states in the 
ground state band are indeed of order unity.  
While this confirms the self-consistency of this symmetric prescription 
of $\lambda$ counting rules, it does not preclude the possible existence 
of other self-consistent counting schemes.  
In this section, we will briefly describe other possible realizations 
of this emergent symmetry.

To gain physical insight, it is useful to think in terms of the bound state 
picture.  
Clearly our simple harmonic oscillator obtained by assuming that $\vec a$, 
$\vec a^\dag \sim \lambda^0$ implies that the ground state expectation value 
$\langle G|x^2|G\rangle\sim \lambda^{1/2}$ vanishes in the combined limit.  
This reflects that, as the reduced mass $\mu\to\infty$, the center of the 
brown muck gets more and more localized around the heavy quark.  
This is the scenario where the origin of the relative position space 
$\vec x=0$ minimizes the potential energy globally.  
However, it does not need to be the case.  
The potential may have a ``mexican hat'' shape with the global minimum at 
$r=|\vec x|=r_0>0$, where by naturalness $r_0\sim 1/\Lambda_{\rm QCD} \sim 
\lambda^0$.  
In such a case, in the combined limit the relative wave function will be 
a shell sharply peaked around $r=r_0$.  
As a result, $\langle G|x^2|G\rangle = r_0^2 \sim \lambda^0$ and the 
symmetric prescription is clearly inapplicable.  
Instead this ``mexican hat'' scenario corresponds to a different realization 
of the emergent symmetry, which hereinafter will be referred to as the 
``symmetry broken realization''.  

We will sketch how one may study the emergent symmetry in this ``mexican hat'' 
scenario, where there are two modes of low-energy excitations.  
Firstly, there are orbital excitations along the bottom of the potential well 
at $r=r_0$, described by the Hamiltonian ${\cal H}_L = L^2/2{\cal I}$.  
The moment of inertia ${\cal I}=\mu r_0^2 \sim \lambda^{-1}$ in the combined 
limit.  
As a result, as $\lambda\to 0$, ${\cal H}_L\to 0$ and the whole tower of 
orbitally excited states collapses into degeneracy.  
The symmetry group of ${\cal H}_L$ with finite moment of inertia $\cal I$ is 
the rotational group O(3), and the spectrum generating algebra is contracted 
O(4) (also known as $E_3$, the three dimensional Euclidean group;  
cf.~Sec.~III.4 of Ref.~\cite{AB}).  
Secondly, there are radial excitation around $r=r_0$, which can be studied 
through a formalism similar to what we constructed in previous sections to 
study the symmetric realization. 
It turns out that the radial excitations are also simple harmonic in the 
combined limit, but only as in a one-dimensional oscillator again with 
$\omega\sim \lambda^{1/2}$.  
The spectrum generating algebra in this case is again a contracted O(4) 
(generated by $a$, $a^\dag$, $a^2$, ${a^\dag}^2$, $a^\dag a$ and {\bf 1}).  
Near the combined limit where $\lambda$ is small, the rotational excitation 
energies $\sim 1/2{\cal I} \sim \lambda$ are much smaller than the radial 
excitation energies $\sim \omega\sim \lambda^{1/2}$.  
Hence the spectrum consists of a tower of equally spaced simple harmonic 
levels with splitting $\omega$, with each level further split into a tower 
of rotor states with splitting $\sim 1/2{\cal I}$.  
As $\lambda\to 0$, all these orbitally and radially excited states become 
degenerate with the ground state, and the spectrum generating algebra 
contracted O(4) $\times$ O(4) becomes the emergent symmetry group in this 
symmetry broken realization.  

This contracted O(4) $\times$ O(4) group in the symmetry broken realization is 
a subgroup of the contracted O(8) in the symmetric realization.  
Actually this is very reminiscent of spontaneous symmetry breaking in field 
theory.  
Note that orbital excitations are light in the combined limit. 
As a result, one can construct a wave function sharply peaked at $x=y=0$, 
$z=r_0$ which is degenerate with the ground state (and hence is itself also a 
legitimate ground state) as $\lambda\to 0$.  
Such a ground state breaks the contracted O(8) in the symmetric realization to 
the contracted O(4) $\times$ O(4) in the asymmetric realization.  
This explains the terminologies, ``symmetric'' and ``symmetry broken'' 
realizations.  

While this symmetry broken realization of the emergent symmetry may not be as 
aesthetically appealing as the symmetric counterpart, we emphasize that it is 
a viable logical possibility and there is no theoretical justification of 
{\it a priori\/} rejecting this possibility. 
However, these two realizations are phenomenologically distinguishable, as 
least when $\lambda\to 0$.  
In the symmetric realization the excitation energy of the second excited state 
is twice that of the first excited state, where in the symmetry broken 
realization the ratio is 3.  
In the real world, the first excited charmed baryon is around 330 MeV heavier 
than the ground state and about 200 MeV beneath the D-N dissociation 
threshold.  
If future experiments find the second excited charmed baryon beneath this 
dissociation threshold, one would be very tempted to rule out the asymmetric 
realization.  \footnote{
To make such a conclusion, however, one has to check if the corrections 
higher order in $\lambda$ are small.  
Unfortunately, such corrections are likely to be substantial.}  

In this section, we have compared the two possible realizations of the 
emergent symmetry: the symmetric realization when the potential is globally 
minimized at the origin, and the symmetry broken realization when the global 
minimum of the potential is away from the origin.  
(Actually it is logically possible that there is more than one global minimum  
--- one at the origin, while the other is not.  
This scenario, however, is so extremely unnatural and requires such 
fine-tuning that we will not consider it further in this paper.)  
While the preceding discussion relied on the bound state picture, one can 
rephrase it in QCD language in a manner analogous to our analysis of the 
symmetric realization represented in previous sections.  
The symmetry broken realization has the interesting feature of light states of 
excitation energies $\sim \lambda$, which originate from the orbital 
revolution around the bottom of the ``mexican hat'' potential.  
But let us be reminded that there are other possible rotational modes for a 
baryon which has nothing to do with orbital revolution.  
For example, one can rotate the brown muck itself (not around the heavy quark) 
in space or in isospace, which quantum mechanically correspond to spin and 
isospin excitations.  
The moment of inertia is of order $N_c$, which leads to a whole tower of rotor 
states which are degenerate in the large $N_c$ limit.  
This is the well-known large $N_c$ spin-flavor symmetry 
\cite{LN3,LN4,LN5,LN6,LN7}, relating $\Delta(1232)$ to the nucleon and 
$\Sigma_Q^{(*)}$ to $\Lambda_Q$.  
These rotational modes of the brown muck have little to do with its relative 
motion relative to the heavy quark.  
As a result, intuitively we expect this large $N_c$ spin-flavor symmetry to 
commute with the emergent symmetry (in either realization).  
It will be the goal of the next section to demonstrate that this intuition is 
indeed correct.  

\section{Inclusion of the spin and isospin effects}

So far we have neglected the spin and isospin of the heavy baryon, which 
consists of a single heavy quark and $N_c-1$ valence light quarks.  
For concreteness we will only consider the cases where $N_c$ is an odd number, 
so there is an even number of valence light quarks in a heavy baryon.  
In QCD with two light flavors, each light quark is isospin-$1\over 2$, and 
as a result the brown muck can be of isospin $I=0$, 1, \dots $(N_c-1)/2$.  
The isospin symmetry is described by an SU(2)$_I$ group, generated by 
\begin{equation}
I^a = \int d^3x \; \sum_{k=1}^{N_c-1} q_k^\dag \tau^a q_k, \qquad a=1, 2, 3, 
\end{equation}
where the summation is over all the valence light quarks.  
Similarly, light quarks are spin-$1\over2$ fermions; a brown muck with $N_c-1$ 
light quarks without any orbital angular momentum between them can be of spin 
$S_\ell=0$, 1, \dots $(N_c-1)/2$.  
(Note that in the heavy quark limit the heavy quark spin $S_Q$ decouples 
from the rest of the system.  
As a result, the brown muck spin $S_\ell$ is conserved and is a good quantum 
number.)  
The brown muck spin symmetry is also described by an SU(2)$_{S_\ell}$ group, 
generated by 
\begin{equation}
S_\ell^i = \int d^3x \; \sum_{k=1}^{N_c-1} q_k^\dag \sigma^i q_k, \qquad 
i=1, 2, 3.  
\end{equation}
Both isospin and brown muck spin symmetries are symmetries of the QCD 
Lagrangian (the latter only in the heavy quark limit), and their generators 
$I^a$ and $S_\ell^i$ satisfy these commutation relations: 
\begin{equation}
[I^a,I^b]=i \epsilon^{abc} I^c, \quad 
[S_\ell^i,S_\ell^j]=i \epsilon^{ijk} S_\ell^k, \quad
[I^a,S_\ell^i]=0.  
\end{equation}

It was realized in 1993 by several different collaborations 
\cite{LN5,LN6,LN7} that for large $N_c$ baryons, the separate isospin and 
brown quark spin symmetries, described by the product group 
SU(2)$_I \times$ SU(2)$_{S_\ell}$, get combined and enlarged into 
an emergent spin-flavor symmetry.  
This spin-flavor symmetry is described by a contracted SU(4) group, generated 
by $\{X^{ai}, I^a, S_\ell^i\}$, satisfying the following commutation 
relations: 
\begin{equation}
[X^{ai},I^b]=i \epsilon^{abc} X^{ci}, \quad 
[X^{ai},S_\ell^j]=i \epsilon^{ijk} X^{ak}, \quad
[X^{ai},X^{bj}]=0.
\end{equation}
with $X^{ai}$ being the axial current couplings: \footnote{
Do not confuse the axial current couplings $X^{ai}$ with $X_j$, the 
center-of-mass position!} 
\begin{equation}
X^{ai} = \int d^3x \; \sum_{k=1}^{N_c-1} q_k^\dag \tau^a \sigma^i q_k, \qquad 
a, i=1, 2, 3.  
\end{equation}
Since all generators commute with $K^2$, where $K$ is the vector sum of 
$I$ and $S_\ell$, all the states with the same $K^2$ are degenerate. 
Of particular interest are the $K=0$ states, for which $I=S_\ell$ and hence 
$(I,S_\ell) = (0,0)$, (1,1), \dots, $((N_c-1)/2,(N_c-1)/2)$.  
Phenomenologically one can identify the (0,0) state as $\Lambda_Q$ and the 
(1,1) state as $\Sigma^{(*)}_Q$, and the $\Sigma^{(*)}_Q$-$\Lambda_Q$ 
splitting is $1/N_c$ suppressed.  
These analyses have been extended to orbitally excited baryons in 
Refs.~\cite{PY1,PY2,JL1,JL2}, although orbitally excited baryons containing 
charm or bottom quarks were only briefly discussed.  

Note that $X^{ai}$, $I^a$ and $S_\ell^i$ act only on the light quarks.  
In other words, they represent the internal degrees of freedom of the brown 
muck, while leaving the heavy quark alone.  
On the other hand, the creation and annihilation operators which generate 
the contracted O(8) symmetry represent collective excitations; 
{\it i.e.}, all the light quarks are excited in a correlated manner relative 
to the heavy quark.  
Intuitively, one expects these two modes of excitations to be independent of 
each other in the large $N_c$ limit, and hence the light quark spin-flavor 
symmetry group should commute with the contracted O(8) group.  \footnote{
We note in passing that models based on the bound state picture have 
often implicitly assumed that spin-flavor degrees of freedom commute with the 
orbital ones.  
In the simple bound state model \cite{genius1,genius2,genius3}, for example, 
this assumption was built in by using the same profile function to describe 
the light baryon as a chiral soliton.}  

To show that this piece of intuition is correct, we will prove the following 
general result: 
Let $J_\ell$ be a {\it local\/} operator which acts only on the brown muck, 
{\it i.e.}, $[{X_Q}_j,J_\ell] = [{P_Q}_j,J_\ell] = 0$.  
Then the forward matrix element of $[a_j,J_\ell]$ and $[a^\dag_j,J_\ell]$ all 
vanish up to order $\lambda$.  
For forward matrix element we mean the kinematic condition that there exists 
an inertial frame such that both the initial and final states are at rest 
(carrying zero total momentum).   

To see that this is true, note that 
\begin{equation}
[a_j,J_\ell] = \sqrt{\mu\omega\over 2} [x_j,J_\ell] + 
i \sqrt{1\over 2\mu\omega} [p_j,J_\ell], \qquad 
[a^\dag_j,J_\ell] = \sqrt{\mu\omega\over 2} [x_j,J_\ell] - 
i \sqrt{1\over 2\mu\omega} [p_j,J_\ell]
\label{ac}
\end{equation}
and the vanishing of the $\vec x$ and $\vec p$ commutators on the right-hand 
side imply the vanishing of the $\vec a$ and $\vec a^\dag$ commutators on the 
left-hand side.  
It is straightforward to show from the definitions of the kinematic 
variables that 
\begin{equation}
\vec x = {1\over \hat\alpha} (\vec X - \vec X_Q), \qquad 
\vec p = \hat\beta \vec P - \vec P_Q, 
\end{equation}
where $\hat\alpha$ and $\hat\beta$ are the operator-valued coefficients 
introduced in Eq.~(\ref{ab}).  
Note that, since $\hat\alpha$ is a $c$-number up to order $\lambda$, 
$1/\hat\alpha$ is well defined up to the same order.  
Since $[{X_Q}_j,J_\ell] = [{P_Q}_j,J_\ell] = 0$, one has 
\begin{equation}
[x_j,J_\ell] = {1\over \hat\alpha} [X_j,J_\ell], \qquad 
[p_j,J_\ell] = \hat\beta [P_j,J_\ell].  
\end{equation}
We have expressed the commutators of the relative kinematic variables 
$\vec x$ and $\vec p$ in terms of their center-of-mass counterparts $\vec X$ 
and $\vec P$. 
However, since $J_\ell$ is a local operator without any intrinsic momentum 
scale, $[X_j,J_\ell]=0$.  
On the other hand, $[P_j,J_\ell]$ in general does {\it not\/} vanish as the 
local operator is translated in position space by $\vec P$.  
The forward matrix element, however, trivially vanishes.  
As a result, both terms in Eqs.~(\ref{ac}) are zero, and the proof is 
completed.  

Two comments are in place here.  
First, by saying that the commutators $[a_j,J_\ell]$ and $[a^\dag_j,J_\ell]$ 
vanish, we actually mean the more precise statement that these commutators 
are smaller than the typical matrix element of $J_\ell$ by at least an order 
in $\lambda$ --- the order at which it is no longer justifiable to treat 
$\hat\alpha$ and $\hat\beta$ as $c$-numbers.  
Second, by reversing the role of the brown muck and the heavy quark, it is 
straightforward to show that, for a {\it local\/} operator $J_Q$ which 
acts only on the heavy quark, 
{\it i.e.}, $[{X_\ell}_j,J_Q] = [{P_\ell}_j,J_Q] = 0$, the forward 
matrix element of $[a_j,J_Q]$ and $[a^\dag_j,J_Q]$ also vanish up to 
order $\lambda$.  
Both of these results reflect the intuitive statement that any excitation 
which involves only one of the constituents but does not transfer any 
momentum will not change the relative motion.  
These results will be useful when we consider the heavy baryon 
matrix elements of pionic current or weak interaction currents in the 
companion paper \cite{next}.  

\bigskip

Returning to our discussion of spin and isospin effects, since the spin-flavor 
symmetry generators $X^{ai}$ act only on the light degrees of freedom, they 
commute with the creation and annihilation operators implying that all of the 
states below are degenerate in the combined limit.  
The states are labeled by the quantum numbers $(N,L,S_\ell,J_\ell)$, where 
$N$ is the number of excitation quanta in the simple harmonic oscillator, $L$ 
is the orbital angular momentum, $S_\ell$ is the spin of the brown muck 
(which is always equal to the isospin $I$ for the $K=0$ states which we are 
working on).  
$J_\ell = S_\ell + L$ is the {\it total\/} (including both orbital and spin) 
angular momentum of the brown muck.  
The total spin of the whole heavy baryon $J=S_Q+J_\ell$ is the vectorial sum 
of $J_\ell$ with the heavy quark spin $S_Q$.  \footnote{
Here we have many different angular momenta adding in different ways.  
We will clarify their meanings by comparing a heavy baryon to a 
multi-electronic atom.  
If the heavy quark is the analogy of the heavy nucleus and the electron 
cloud corresponds to the brown muck, then $S_\ell$ corresponds to the 
electronic spin $\cal S$, $L$ corresponds to the orbital angular momentum 
$\cal L$, and $J_\ell = S_\ell + L$ translates into ${\cal J}={\cal S}+
{\cal L}$.  
The heavy quark spin $S_Q$ is the counterpart of the nuclear spin ${\cal I}$, 
and the total spin of the heavy baryon given by $J=S_Q+J_\ell$ is the 
analogy of the $\cal F$-spin, ${\cal F}={\cal I}+{\cal J}$. }  

\begin{equation}
\matrix{
& & &\vdots& & &\vdots& &\cr
& & &\Bigg\uparrow a_j^\dag& & &\Bigg\uparrow a_j^\dag& & \cr
N=2&L=0,2& &\Lambda^{(*)}_{Q2}&(J_\ell=0,2)&\buildrel X^{ai}\over
{\hbox to 40pt{\rightarrowfill}}&\Sigma^{(*)}_{Q2}&(J_\ell=1,2,3)&\buildrel 
X^{ai}\over{\hbox to 40pt{\rightarrowfill}}&\dots\cr
& & &\Bigg\uparrow a_j^\dag& & &\Bigg\uparrow a_j^\dag& & \cr
N=1&L=1  & &\Lambda^{(*)}_{Q1}&(J_\ell=1)&\buildrel X^{ai}\over
{\hbox to 40pt{\rightarrowfill}}&\Sigma^{(*)}_{Q1}&(J_\ell=0,1,2)&\buildrel 
X^{ai}\over{\hbox to 40pt{\rightarrowfill}}&\dots\cr
& & &\Bigg\uparrow a_j^\dag& & &\Bigg\uparrow a_j^\dag& & \cr
N=0&L=0  & &\Lambda_Q&(J_\ell=0)&\buildrel X^{ai}\over
{\hbox to 40pt{\rightarrowfill}}&\Sigma^{(*)}_Q&(J_\ell=1)&\buildrel 
X^{ai}\over{\hbox to 40pt{\rightarrowfill}}
&\dots\cr
& & & & & & & & \cr
& & &\hbox{The $\Lambda$ sector}& & &\hbox{The $\Sigma$ sector}& &\cr
& & &I=S_\ell=0& & &I=S_\ell=1& &\cr}
\end{equation}
That $X^{ai}$ and $a^\dag_j$ commute means that one can construct, say, 
the $\Sigma^{(*)}_{Q1}$ state through $a^\dag_j (X^{ai} \Lambda_Q)=a^\dag 
\Sigma^{(*)}$, or the opposite order $X^{ai} (a^\dag_j \Lambda_Q)= X^{ai} 
\Lambda^{(*)}_{Q1}$, and both constructions give the same state 
$\Sigma^{(*)}_{Q1}$.  

Recall that we have shown in Eq.~(\ref{result}) that the QCD Hamiltonian in 
the heavy baryon sector can be written as the sum of the simple harmonic 
part ${\cal H}_{\rm SHO}$, and the internal excitation Hamiltonian 
${\cal H}_{\rm exc}$, with possible corrections of order $\lambda$. \footnote{
We are working with the symmetric realization here.  
The case for asymmetric realization can be studied in a similar way.}  
Among the different contributions to ${\cal H}_{\rm exc}$ is 
${\cal H}_I = \sigma I^2$, the Hamiltonian describing the low-lying 
spin-flavor excitations.  
The large $N_c$ spin-flavor symmetry implies $\sigma \sim N_c^{-1} \sim 
\lambda$.  
As a result, ${\cal H}_I\sim\lambda$ and one can move it from 
${\cal H}_{\rm exc}$ to ${\cal H}_{\rm SHO}$ and rewrite Eq.~(\ref{result}) 
as follows: 
\begin{equation}
{\cal H} = {\cal H}'_{\rm SHO} + {\cal H}'_{\rm exc} + {\cal O}(\lambda), 
\quad {\cal H}'_{\rm SHO} = {\cal H}_{\rm SHO} + {\cal H}_I, 
\end{equation}
and ${\cal H}'_{\rm exc}$ is the Hamiltonian for other internal excitation 
modes excluding the low-lying spin-flavor excitations.  

The Hamiltonian ${\cal H}'_{\rm SHO}$ describes the low-lying heavy baryon 
spectroscopy up to order $\lambda$ corrections.  
(Note that these corrections are of the same order as ${\cal H}_I$.) 
Under ${\cal H}'_{\rm SHO}$, each of the simple harmonic bound states of 
${\cal H}_{\rm SHO}$ is split by ${\cal H}_I$ into a whole 
tower of states with different $I$.  
The masses of the heavy baryon states under ${\cal H}'_{\rm SHO}$ are  
\begin{equation}
m = \Lambda_Q + N\omega + \sigma I^2 + {\cal O}(\lambda), 
\end{equation}
where $N=n_x + n_y + n_z$ is the total number of excitation quanta, and we are 
adopting the customary abuse of notation that the symbol of a state also 
represent its mass, {\it e.g.}, $\Lambda_b=m_{\Lambda_b}=5624$ MeV.  
This mass relation implies that the orbital excitation energies are the same 
in the $\Lambda$ sector ($I=0$) and the $\Sigma$ sector ($I=1$), {\it i.e.}, 
\begin{equation}
\omega_\Sigma - \omega_\Lambda \equiv (\Sigma^{(*)}_{Q1} - \Sigma^{(*)}_Q) 
- (\Lambda^{(*)}_{Q1} - \Lambda_Q) = {\cal O}(\lambda).  
\end{equation}
Formally it also implies spin-flavor excitation energies of the $N=1$ states 
are the same as their $N=0$ counterparts: 
\begin{equation}
2\sigma_1 - 2\sigma_0 \equiv (\Sigma^{(*)}_{Q1} - \Lambda^{(*)}_{Q1}) 
- (\Sigma^{(*)}_Q - \Lambda_Q) = {\cal O}(\lambda).  
\end{equation}
Unfortunately this relation is actually devoid of information, as both 
$\sigma_1$ and $\sigma_0$ are of order $\lambda$, which is the order of the 
leading order corrections.  

Note that the qualitative features of the low-lying heavy baryon spectrum is 
correctly given by ${\cal H}'_{\rm SHO}$, which specifies the low-energy 
spectroscopy up to order $\lambda^{1/2}$.  
The not-yet-specified correction terms at order $\lambda$ are small when 
compared to ${\cal H}'_{\rm SHO}$ and hence can only perturb the spectrum 
but not change it qualitatively.  
On the other hand, to make quantitative predictions about heavy baryon 
spectroscopy (or other heavy baryon dynamical properties) at order 
$\lambda$, one has to determine the explicit forms of the order $\lambda$ 
correction terms --- a task which we will undertake in our next paper 
\cite{next}.  

\section{concluding remarks}

In this paper, we have shown that there is an emergent symmetry in the 
heavy baryon Hilbert subspace in the combined heavy quark and large $N_c$ 
limits.  
This emergent symmetry can either be realized as a contracted O(8) or a 
contracted O(4) $\times$ O(4), and in either case it relates the orbitally 
excited states with the ground state.  
Both realizations of this emergent symmetry have interesting spectroscopic 
implications: in the former case the spectrum is that of a three-dimensional 
simple harmonic oscillator, while in the latter case the spectrum is that of a 
heavy rigid rotor with simple harmonic radial excitations.  
Moreover, we have also shown that this emergent symmetry commutes with the 
light quark spin-flavor symmetry for large $N_c$ baryons.  

While the main purpose of this paper is to discuss the results reported in 
Ref.~\cite{old} in a more detailed fashion, there are several places where 
the formalism in this paper differs from the original treatment in 
Ref.~\cite{old}.  
We list the most important differences below for comparison:  

$\bullet$  In Ref.~\cite{old}, we have always taken the heavy quark limit 
before the large $N_c$ limit.  
In this paper, the combined limit is taken by keeping $N_c \Lambda_{\rm QCD}
/m_Q$ fixed at an arbitrary value.  
This is a more general treatment, as it includes the possibilities of taking 
the heavy quark limit both before and after the large $N_c$ limit, and a 
whole range of other possible limiting procedures. 

$\bullet$  Since the heavy quark limit is taken first in Ref.~\cite{old}, 
one does not need to distinguish the relative momentum $\vec p$ from the brown 
muck momentum $\vec P_\ell$.  
On the other hand, in the paper we are keeping the ratio $N_c \Lambda_{\rm QCD}
/m_Q$ arbitrary, and the distinction between $\vec p$ and $\vec P_\ell$ should 
not be overlooked.  
Here we have presented the analysis of the kinematics in full generality, 
while the result of Ref.~\cite{old} can be recovered by setting $\hat\alpha=0$ 
and $\hat\beta={\bf 1}$.  

$\bullet$  We have realized that the emergent symmetry group reported in 
Ref.~\cite{old}, namely the contracted U(4), is only a subgroup of the full 
symmetry group in the symmetric realization --- contracted O(8).
Moreover, we have studied the symmetry broken realization for the sake of 
generality.  

We want to emphasize once more that, even though we seem to be dealing 
with a quantum mechanical potential model, the formalism is actually 
completely field theoretical, and the kinematic variables are QCD operators.  
Why can one reduce a field theoretical problem into a quantum mechanical 
framework?  
The answer lies in the separation of scales: both constituents, namely 
the heavy quark and the brown muck, have mass of order $\lambda^{-1}$, 
while the interaction is only of order unity and hence is very weak 
compared to the mass scale represented by the reduced mass $\mu$.  
The separation of scales makes it possible to make an expansion in powers 
of $\lambda$ and have an effective field theory which includes only the 
lowest excitation modes, which in this case is the motion of the brown muck 
relative to the heavy quark (and the possible spin-flavor excitations). 
This falls under a category of effective field theory which is referred to as 
``rigorous potential models'' in Ref.~\cite{PL}, of which the most notable 
example is nonrelativistic QCD (NRQCD) \cite{BBL}.  
NRQCD describes quarkonium states, which are heavy quark--heavy antiquark 
bound states.  
While the mass of each constituent is $m_Q$, the three-momentum of the 
relative motion is only of the order $\alpha_s m_Q$, and the kinetic 
energy is of an even lower order $\alpha_s^2 m_Q$, where $\alpha_s$ is the 
QCD coupling constant at scale $m_Q$.  
Since $\alpha_s$ is small, we have a separation of scales which allows us 
to expand the Hamiltonian in powers of $\alpha_s$.  
In particular, since the kinetic energy scale is much smaller than the 
three-momentum scale, it is natural to impose a stronger cutoff on energies 
than on three-momenta in the effective field theory.  
The resultant effective theory is local in time but not in space, {\it i.e.}, 
a potential.  
Potential models constructed with such a philosophy are rigorous in the 
sense that they are related to the original field theory through Wilsonian 
renormalization. 
Our treatment of heavy baryons is similar to NRQCD, with $\lambda^{1/2}$ 
playing the role of $\alpha_s^2$.  
The similarity is even more apparent when one realizes that our formalism, 
just like NRQCD, can be viewed as a nonrelativistic expansion.  
In NRQCD $v^2\sim\alpha_s^2$, while in our case $v^2\sim \lambda^{1/2}$.  

We will end on a cautionary note with a comparison with another ``rigorous 
potential model'', namely the nucleon-nucleon effective field theory.  
The deuteron is a nonrelativistic bound state, with the binding energy 
$\sim 2$ MeV order of magnitude smaller than the nucleon mass $\sim 1$ GeV.  
However, the large $N_c$ counting rules would have suggested a very 
different picture.  
Since baryon-baryon interaction is of order $N_c$, both the binding energy 
$V_0$ and the spring constants $\kappa$ of a deuteron are of order $N_c$.  
With reduced mass $\mu\sim N_c$, large $N_c$ scaling rules suggest that 
the excitation energy $\omega=\sqrt{\kappa/\mu} \sim N_c^0$, which is 
much smaller than the binding energy $V_0$.  
This implies the existence of many bound states beneath the 
dissociation threshold (the number of bound states should be of order 
$V_0/\omega \sim N_c$), and the low-lying bound states should be deeply 
bound.  
This is in blatant disagreement with the deuteron in the real world, which 
is barely bound with a tiny binding energy in comparison to 
$\Lambda_{\rm QCD}$: $V_0\sim 2$ MeV in the triplet channel, and the singlet 
channel is not even bound.  
It turns out that there are many different physical contributions to the 
binding energy.  
Each of these contributions may be of order $N_c$, but it happens that they 
almost cancel completely, and the numerical value for $V_0$ turns out to be 
accidentally small.  
This illustrates a fundamental limitation of counting schemes: a physical 
quantity may carry a {\it numerical\/} value very different from what the 
{\it formal\/} power counting suggests due to accidental cancelation or 
appearances of unnaturally large or small coefficients.  
One should be aware of the possibilities of such 
accidents and carefully check if the physical picture suggested by the 
counting rules resembles the real world.  

For our analysis of the heavy baryon, the binding energy $V_0$ is formally 
of order unity, while the excitation energy $\omega$ is of order 
$\lambda^{1/2}$. 
As a result, one expects the number of bound states to be of order 
$\lambda^{-1/2} \sim N_c^{1/2} = \sqrt{3}$ in the real world.  
Experimentally $V_0$ and $\omega$ have been determined to be about 625 MeV 
and 330 MeV, respectively.  
These numbers are at least compatible with the picture suggested in 
this paper.  
With an expansion parameter as large as $\lambda^{1/2}\sim 1/\sqrt{3} \sim 
0.6$, however, the expansion series may converge slowly and one needs to 
include corrections due to higher order terms before one can make any 
quantitative predictions.  
As a result, it is imperative to make a careful study of the higher order 
corrections.  

In summary, we have introduced an effective theory to study excited heavy 
baryons which makes the existence of the emergent symmetry manifest.  
What are the phenomenological implications of this emergent symmetry?  
What does this symmetry tell us about the strong decays of excited baryons 
and the weak decay form factors?  
And most importantly, are these symmetry predictions safe against corrections 
due to higher order terms in the $\lambda$ counting?  
All of these issues will be discussed in an upcoming paper \cite{next}.  

\bigskip

This work is supported by the U.S.~Department of Energy grant 
DE-FG02-93ER-40762.

\end{document}